\documentclass[]{article}
\pdfoutput=1
\usepackage{cite}
\usepackage{amsmath,amssymb,amsfonts}
\usepackage{algorithmic}
\usepackage{graphicx}
\usepackage{xspace}
\usepackage{multirow}
\usepackage{url}
\usepackage{enumitem}
\usepackage{color}
\usepackage[tight,footnotesize]{subfigure}
\usepackage[usenames,dvipsnames]{xcolor}
\usepackage[ruled,vlined]{algorithm2e}
\usepackage{cleveref}
\usepackage{microtype}
\usepackage{placeins}

\newcommand{\algo}{PCach\xspace}
\newcommand{\app}{MACACO-app\xspace}

\begin{document}

\title{\algo: The Case for Pre-Caching your Mobile Data}

\author{Katia Jaffr\`es-Runser and Gentian Jakllari\\ University of Toulouse -- IRIT\\
Toulouse, France \\
\{kjr, jakllari\}- at - enseeiht.fr
}

\maketitle

\begin{abstract}
We present PCach, a smartphone-based approach for relieving the congestion in cellular 
network resulting from the exponential growth in mobile data traffic.
 The basic idea underlying PCach is simple: use WiFi to
proactively cache content on the smartphone's memory, which otherwise would have
been delivered through the cellular network. However, it leads to several challenging questions,
including how much mobile data actually flows through cellular networks, how much
data can be pre-cached, and when and what to pre-cache. 
We address these questions progressively
using a thorough analysis of user data collected from our purpose-built crowdsensing Android application,
actively utilized by 45 users for periods dating back to July 2014. Our analysis shows that the median
smartphone user transfers 15\% of their data via the cellular network and that 80\% of it 
can be pre-cached via WiFi. To capitalize on these results, we draw on a careful
analysis of the measurement data to introduce an algorithm that can run
stand-alone on off-the-shelf smartphones and predict with good accuracy
when and what to pre-cache.

\end{abstract}


\section{Introduction}

Recently, we marked the 10th anniversary of the launch of the first iPhone, which sparked the 
smartphone revolution and has had a big
impact on how we generate and consume content. It has been a boon for cellular providers but also an extreme challenge  due to the exponential growth in demand for mobile capacity. Cisco, in its widely cited Visual Networking Index~\cite{cisco-vni}, reports
that mobile data traffic has grown 4,000-fold over the past 10 years. This growth is largely due
to the rise of the smart devices. In 2016, 89\% of the mobile data traffic
was generated by smart devices. Consecutive generations of mobile telecommunications technologies (3G, 4G) have followed, with ever expanding
capacities, responding to the exponentially increasing appetite for mobile data. Nevertheless, all indications point to the data volume growing
faster than the mobile capacity. For example, Cisco projects the mobile data traffic to increase nearly eightfold between 2015 and 2020, with smartphones
responsible for four-fifth of the total volume. Mobile network connection speeds, on the other hand,
will only increase by threefold by 2020. 

5G is proposed as the answer to the looming data crunch. Its goal is to offer edge\footnote{Edge rate is the worst rate 95\% of the users can reasonably expect.}
 data rates ranging
between 100 Mbps to 1 Gbps, a 100 to 1000 factor improvement over 4G~\cite{5GBE}. This spectacular improvement
is probably the only thing clear about 5G -- getting there is still subject to intense debate in academia
and industry. Nevertheless, there is general agreement that the increase in performance will be achieved
through a combination of 
Millimeter Wave (mmWave)~\cite{rangan2014,80211ad}, ultra-densification~\cite{6620224,andrews2012}, massive
multiple-input, multiple-output (MIMO)~\cite{marzeta2007,marzeta2010} and edge caching~\cite{Debbah-caching-2014}. All these solutions, 
however, require fundamental changes to the architecture of mobile networks and are still years away. 
Cisco projects that
by 2021, 5G will represent only 0.2 percent of connections and 1.5 percent of total traffic. Solutions that can be deployed immediately 
and  serve as bridge to the 5G roll-out are sorely needed. 


In this paper, we introduce PCach, a user-centric approach that can be deployed on off-the-shelf smartphones by simply downloading
an application and help relieve
congestion in cellular networks. To accomplish this, PCach uses the Wi-Fi connection to proactively cache (pre-cache) content users are likely to need in the immediate future and otherwise would have 
downloaded through the cellular connection. While the principle may sound simple and
intuitive, it faces several questions and challenges. The most basic question is what is the potential of improvement for such 
an approach. Once the potential established, several research challenges need to be addressed:
First, PCach needs to identify what content a particular user is likely to need in the
immediate future and constitutes a good value for ``immediate". 
Second, it needs to identify when a particular user is about to switch from Wi-Fi to cellular and back. Finally, it needs to combine
everything together into a lightweight application that one can download from any of the popular app stores and can help
relieve congestion, a far bigger concern for the telecommunication companies, without adversely impacting the smartphone
user experience, a far bigger concern for the end-user.

In short, we address these challenges progressively, using a customized Android application deployed on a large number of
smartphones across 5 different countries as an ideas laboratory. Drawing on a careful analysis of the measurement data,
we identify the most cost-effective way to deploy PCach and establish bounds on the amount of cellular traffic it can pre-cache.  
Finally, we apply our data-driven approach to tailor solutions to well-known machine learning challenges to PCach.  

\begin{figure}[tb]
\centering
\subfigure[]{\includegraphics[width=2.5in]{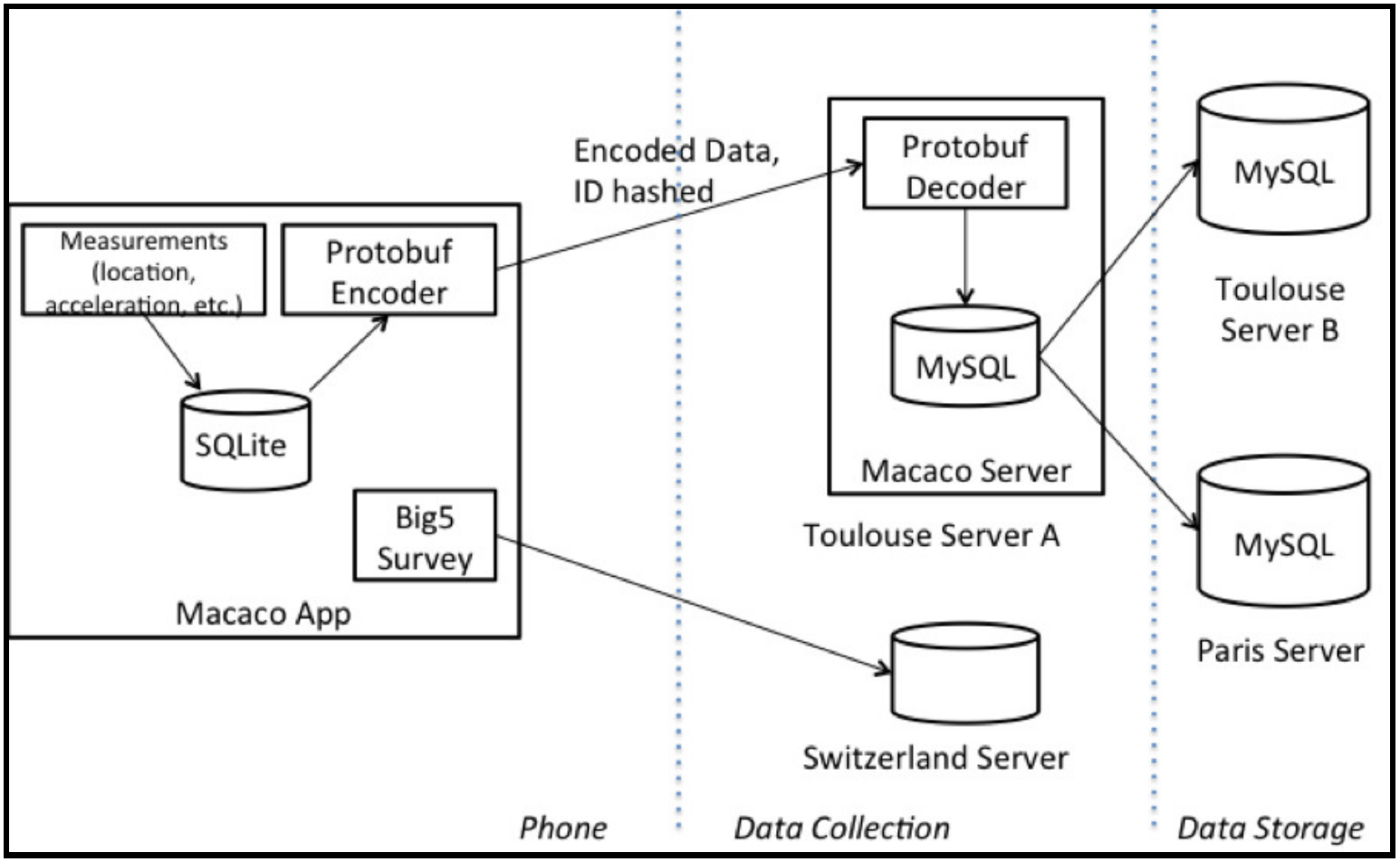}\label{fig:overllarchi}}
\subfigure[]{\includegraphics[width=2.5in]{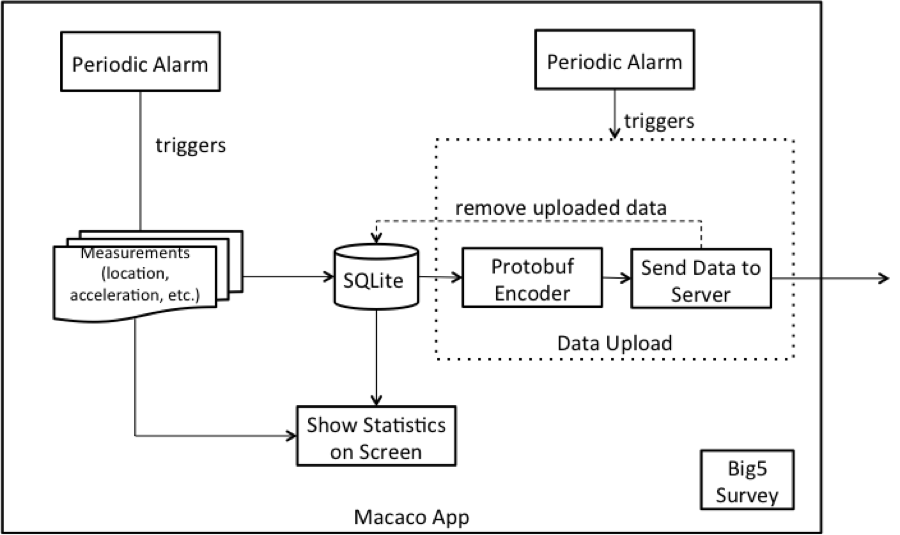}\label{fig:apparchi}}
\caption{The overall crowdsensing architecture (top) and \app architecture (bottom)}
\label{fig:archi}
\end{figure}

Our main contributions may be summarized as follows:

\begin{itemize}[noitemsep,nolistsep,leftmargin=*]
\item We design and implement \app, an Android application for crowdsourcing fine-grained statistics on networking content and context (\cref{sec:sens-app}). Our design
carefully addresses the demands of a successful crowdsourcing application, including user privacy, impact on energy consumption and user incentive.
 \app ha been actively utilized by 45 users from 5 different countries during the period between July 2014 and July 2017, 
 creating a rich dataset.  
\item Using the dataset, we establish a case for pre-caching (\cref{sec:mining}) by showing that $i)$ a significant amount of mobile
traffic is delivered through the cellular network, $ii)$ there are non-trivial gaps in the WiFi connectivity, concentrated 
around commute time, $iii)$ up to 80\% of the data consumed during the WiFi gaps can be pre-cached.
\item We introduce PCach (\cref{sec:pred}), a user-centric approach that can run as an ordinary app on off-the-shelf
smartphones and pre-cache via WiFi content that otherwise would have been delivered through the cellular
network. Using a data-driven design process, we tailor standard machine learning approaches
to addressing the two main challenges facing PCach: predicting WiFi gaps and what content to pre-cache.  
\end{itemize}

\section{Analyzing Smartphone Usage in the Wild}
\label{sec:sens-app}
To make the case for pre-caching data on smartphones, we have designed and deployed a mobile Android application to crowdsource fine-grained statistics on networking content and context.
This application has been deployed on a set of 45 smartphones for an extended period of time, offering a rich dataset on which we base our analysis of the benefits and design challenges of mobile data pre-caching.  

\subsection{Design of \app, a crowdsensing mobile app}
\label{subsec:archi}

As for any crowdsensing app, the quality and quantity of data that can be collected by \app is strongly conditioned by the motivation of the sensing participants~\cite{crowdsourcing}. We have identified the following 
key conditions for having a sensing app adopted for a long time by participants:
\begin{itemize}
\item the sensing app should not disrupt the participants' experience with their smartphones,
\item it should not put too much stress on the battery,  
\item upload data to the collection server(s) with no impact on cellular traffic,
\item data collection should  respect the participants' privacy,
\item participants should be provided some kind of incentive.
\end{itemize}

\subsubsection{{\bf The App architecture}}

Fig.~\ref{fig:overllarchi} shows the overall architecture of  our data sensing system. It consists of a mobile application, \app, and a system for collecting and storing the sensing data.

\app, whose architecture is shown in Fig.~\ref{fig:apparchi}, runs as a foreground Android service. It implements two periodic alarms, one for triggering the data collection and the other 
for pushing data to the front-end servers. The data collection period is set to 5 minutes. As an incentive mechanism, \app stores the collected data temporarily in a SQLlite database and analyzes it in order to show
 the participants useful statistic as to their daily smartphone usage.

The collected data is uploaded to front-end servers. The one located in Toulouse collects the periodic samples of all 
statistics shown in Table~\ref{tab:data} and stores it in a MySql database. The data collected each day is sent overnight to two storage servers located in Toulouse and Paris.
\app sends the data using an energy efficient serialization library provided by Android. 

To provide the best privacy practices to our participants, we have followed the privacy enforcement rules of CNIL\footnote{CommissionCommission Nationale de l'Informatique et des Libert\'es.}, the French privacy regulation body. 
Thus, all data sent by  the \app users are identified on our server with a SHA-256 hash of the mobile IMEI (International Mobile Equipment Identity). 
Servers use secure communications and access to the stored data is available only to  well-identified project members.  
Finally, data is stored in non-anonymized format for a limited duration only.

\subsubsection{{\bf Energy efficiency}}
\label{sec:energy}
A significant challenged faced by any  crowdsensing app design is how to minimize energy consumption. It is possible to measure the app's energy profile using specific monitoring hardware and software \cite{Alan}. 
Using such a profiling platform, we have found that for \app the highest energy consumer tasks are GPS localization [$\sim$6Wh], followed by Bluetooth, accelerometer, and Wi-Fi scanning [$\sim$0.4Wh each]. 
In the following, we present the solutions we have implemented to reduce the energy footprint of \app. 
\begin{figure}[tb]
\centering
\includegraphics[scale=0.3]{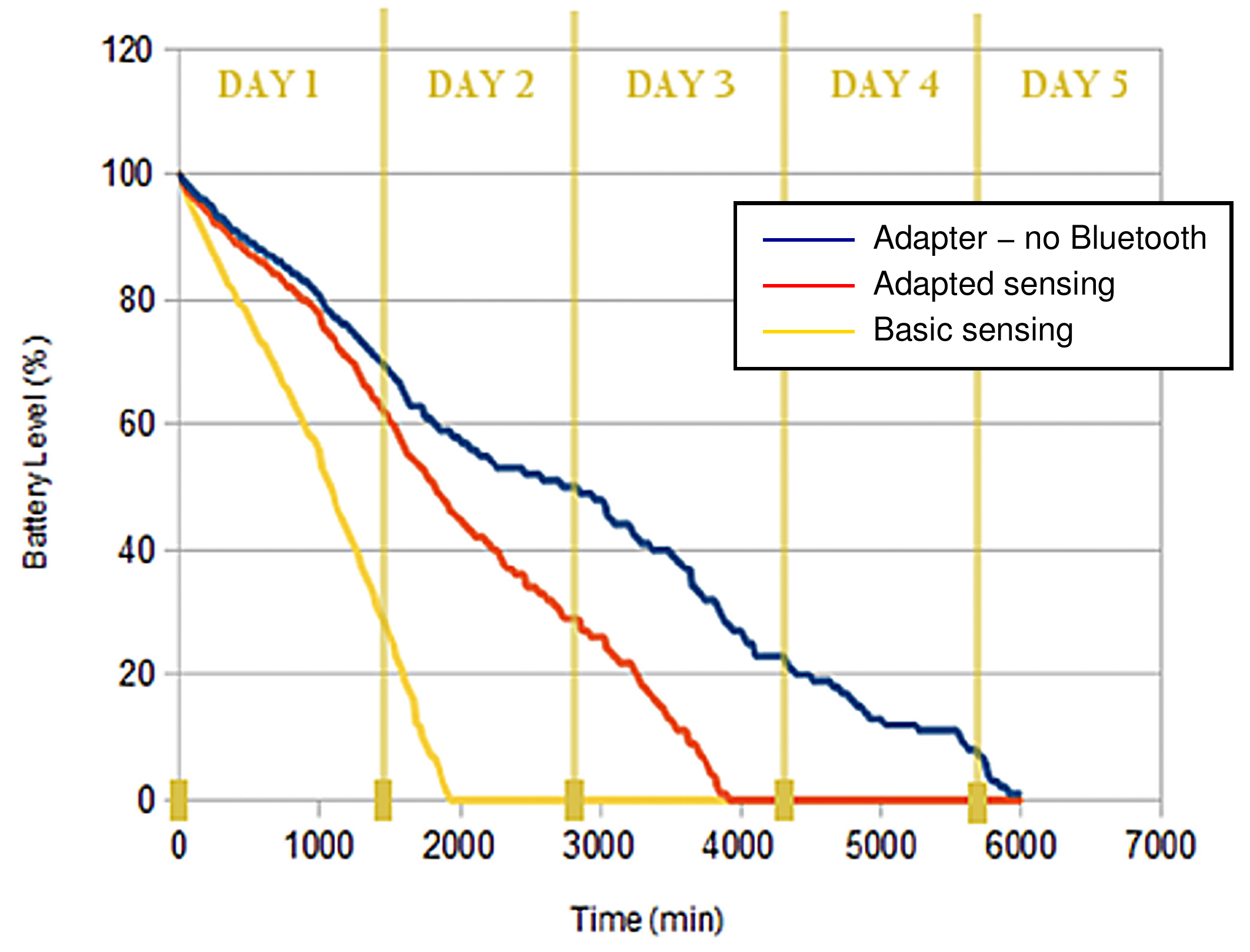}
\caption{Impact of energy-related optimization on context data retrieval}\label{fig:energy}
\end{figure}

\paragraph*{Network sensing} 
WiFi and Bluetooth sensing are energy hungry tasks because the physical layer has to scan all channels for beacons and retrieve all information on available networks. 
Therefore, instead of asking for a new scan at every data collection period, \app polls the Android API to get the lasted scan results, populated periodically by Android's networking interface.   

\paragraph*{Location sensing} The most precise location sensing is obtained by making a GPS service call. However, not only is GPS the most energy-hungry task
but it may fail if not enough satellites are visible (this is often the case indoors).
As a result, a naive data sampling method consisting of simply calling GPS and collecting its result every collection period would be energy hungry and unreliable.

A better solution can be built by better understanding how Android provides localization. It uses a global variable, the \emph{last known location}, where
it stores its best estimate  of the user location and makes it accessible to any app requesting it. To update the value of the \emph{last known location} variable
it uses a combination if its in-house localization protocol and opportunistically collecting GPS data whenever a third application, say a navigation app, asks 
for it.  \app takes advantage of this Android mechanism for location sensing. At every data sampling period, it returns the value of the \emph{last known location} global variable. 
To make sure that it is getting an accurate value, it performs the following trick. First, it checks whether the user has changed location since the last data collection. 
If no, it considers the current value of last known location as precise and takes no further action.
If the user has changed location, it explicitly calls the GPS, triggering Android into collecting the data, if the GPS does return location information, and updating the last known location variable.
To decide whether the user has moved, \app employs a simple heuristic: If the identities of the three strongest access points returned by the WiFi scan have changed so has the user
location, otherwise it has not.      

\paragraph*{Evaluation} Fig.~\ref{fig:energy} shows the energy consumption of a single device (MotoG 1st generation) left stationary.
The battery lifetime is measured when $(i)$ GPS is triggered every 5 minutes (basic), $(ii)$ GPS is triggered only when 
location change is detected (adapted sensing) and when $(iii)$ adapted sensing is coupled with disabling Bluetooth scan. The data shows
that \app's approach extends battery life by 100\% when compared to a straightforward approach. Moreover, disabling Bluetooth improves the battery lifetime by an entire 24 hours.

\subsection{Statistics Collected}
\label{subsec:statscollected}

\app collects the information listed in Table~\ref{tab:data} every 5 minutes, grouped into {\em context and content} features.   
\begin{table}[t]
\centering
\caption{Content and context samples sensed.}\label{tab:data}
\begin{tabular}{|c||c|c|c|}
\hline
\multirow{2}{*}{Context} & WiFi scan & 3G scan      & Bluetooth scan    \\ \cline{2-4} 
                         & Location  & Acceleration & Battery level \\ \hline
\multirow{2}{*}{Content} & URL       & Running apps list  & App upload (bytes)    \\  \cline{4-4}
                         &           &          & App download (bytes) \\  \hline
\end{tabular}
\end{table}

Context defines in our case the mobile user's environment. It can 
be described by various information features, including user location, user motion or specifics on the currently available 
wireless networks. Data about the context features can be  obtained by triggering system calls to various sensors 
(e.g.~GPS, accelerometer, gyroscope, battery level, etc.) or to network interfaces (e.g.~WiFi, Bluetooth, Cellular, etc.). For this study,
 we were particularly interested in the following context features:
\begin{itemize}
\item WiFi scan: the list of visible WiFi networks,
\item whether the smartphone's active network is WiFi or cellular, 
\item if connected to WiFi, the name of the WiFi network the user is currently connected to.
\end{itemize}
\begin{figure}[t]
\centering
\includegraphics[scale=0.4]{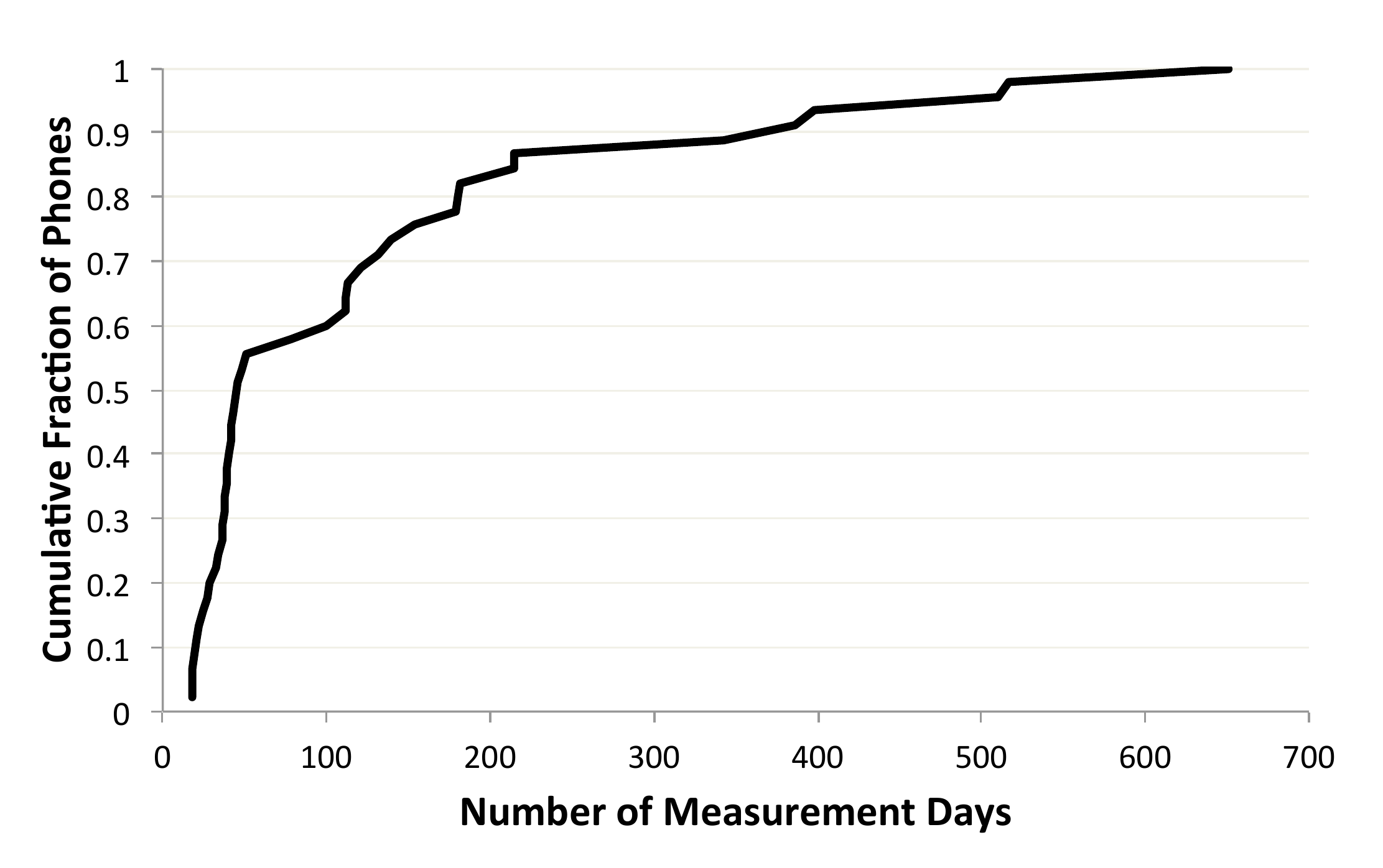}
\caption{Distribution of collection period lengths (in days) for the 45 phones of interest.}
\label{fig:numdays}
\end{figure}

\begin{figure}[htb]
\centering
\includegraphics[width=3in]{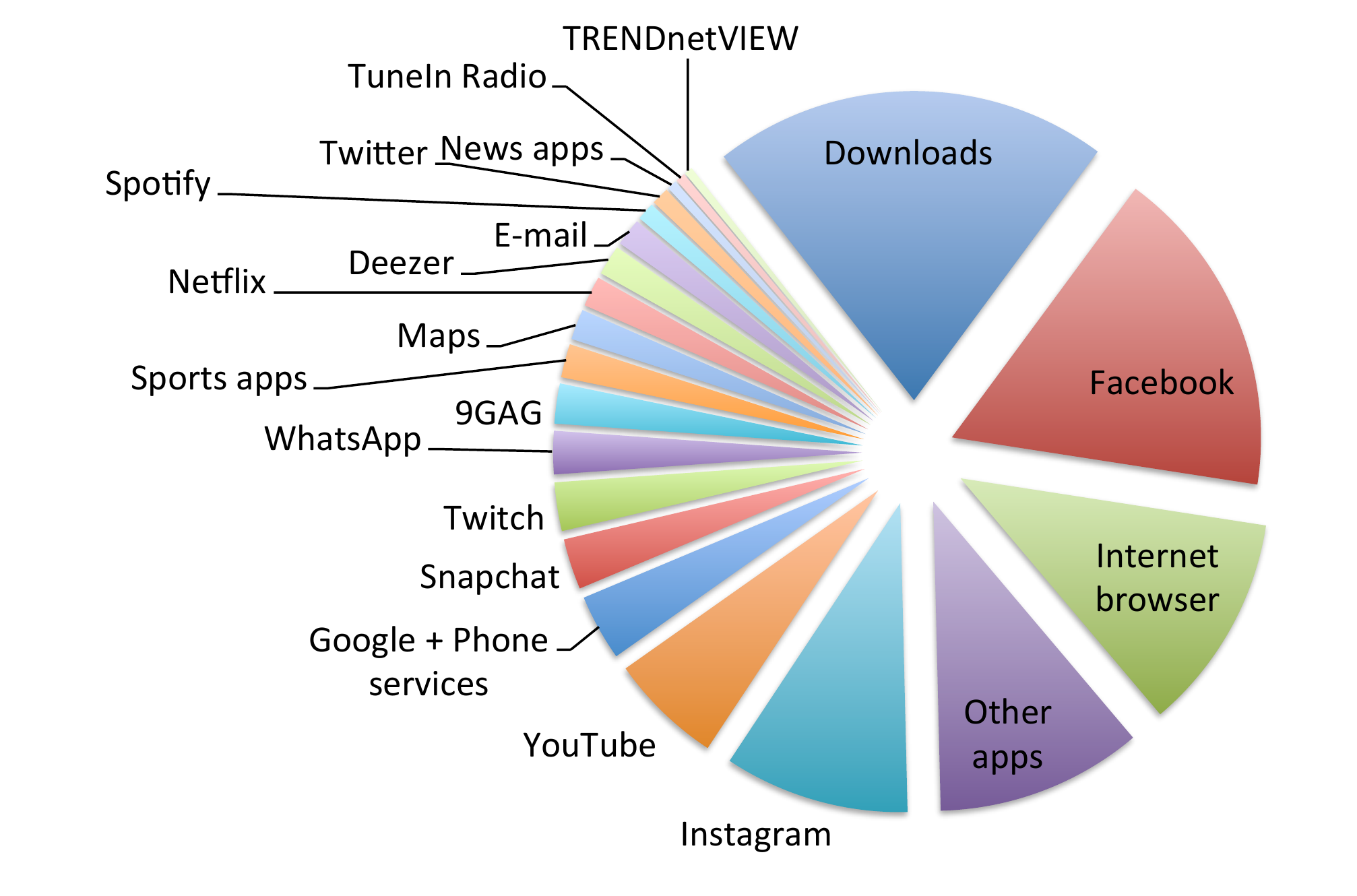}\hfill
\includegraphics[width=3in]{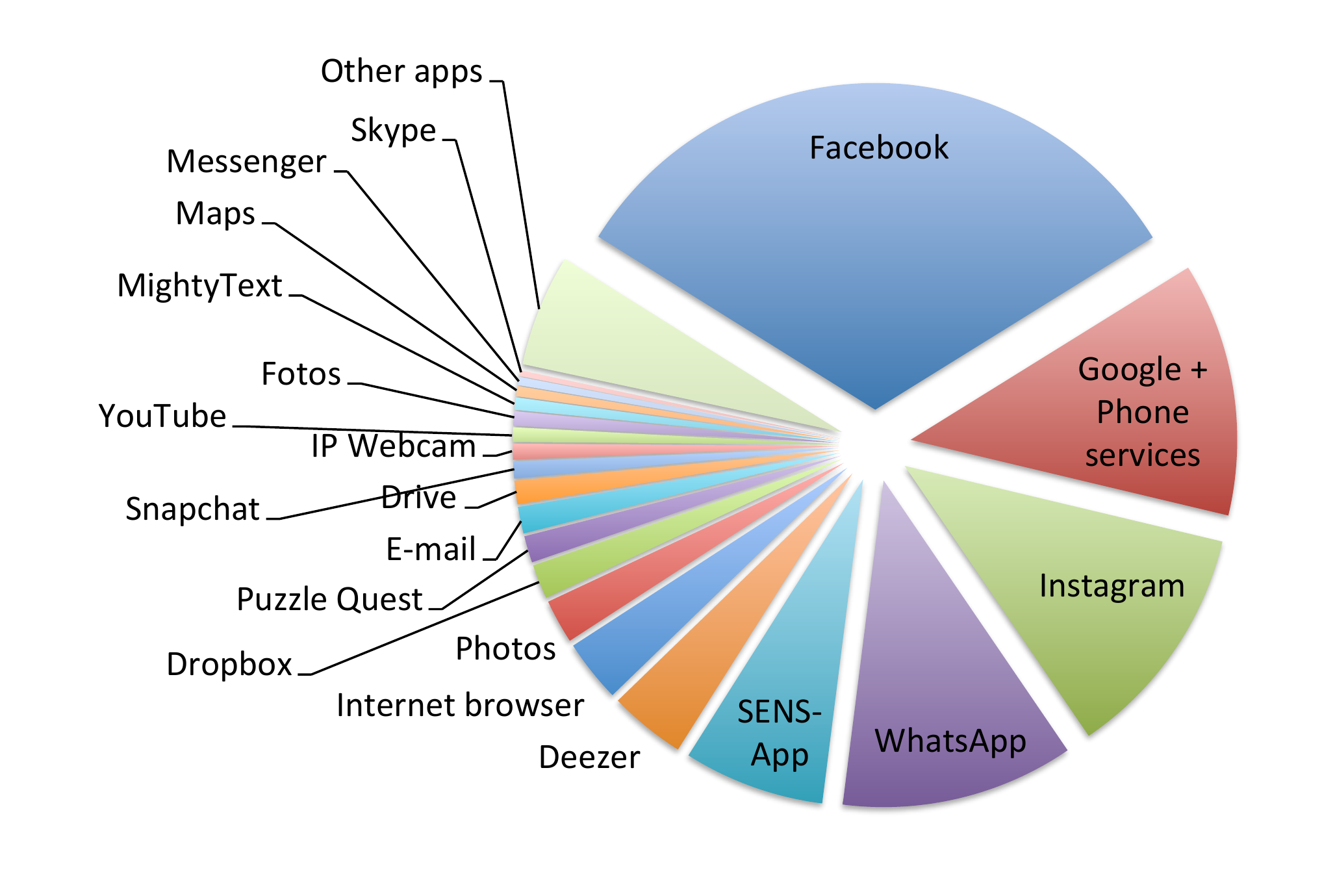}\hfill
\caption{Top 20 applications in terms of download (top) and upload (bottom) data.}
\label{fig:pieApps}
\end{figure}

Content relates to the nature of the data that is either pushed onto or pulled from the Internet by the smartphone user.
It can be described by more or less precise pieces of information, ranging from the accurate description of the data 
(e.g. URL of the Super Bowl 51 highlights video) to the raw information where one can only know that content 
originates from a given application. Accessing content on mobile devices is not straightforward as it depends on the privacy rules enforced by the operating system. 
Most content is manipulated inside mobile applications that have to explicitly grant access to a third party application for monitoring it. 
As we do not want to break the Android's privacy rules, we capture for each measurement the following content features:
\begin{itemize}
\item the list of applications currently running,
\item the volume of data uploaded per application to the Internet, 
\item and the volume of data downloaded per application.
\end{itemize}

\subsection{Dataset}
\label{subsec:dataset}

\app has been installed on 162 smartphones during the period between July 2014 and July 2017, creating a dataset of 2.64 million measurement samples, representing 220k hours of measurements. 
Volunteers originate from 5 different countries, spanning 2 continents. They represent different profiles, including students, full-time employees in academia and industry. 

Fig.~\ref{fig:pieApps} plots a pie chart of the top 20 applications in terms of download volume (top) and upload volume (bottom). To create this chart, we have grouped some similar apps into the categories of Table~\ref{tab:categories}.

\begin{table}[tb]
\centering
\caption{Description of application categories used in Fig.~\ref{fig:pieApps}}
\label{tab:categories}
\begin{tabular}{ll}
Download                & Apps that handle file downloads        \\
Internet browser        & e.g Firefox, Chrome, etc.                                    \\
E-mail apps             & e.g. Gmail, E-mail, Yahoo mail, etc.                         \\
Google + Phone services & Periodic Google, OS and manufacturer services \\
Sports                  & Related to news in sports                                    \\
News                    & Related to news                                               \\
Other apps              & Remaining apps (1392 out of 1412)                                     \\                                             
\end{tabular}
\end{table}

The last category, Other apps, merges the contribution of the remaining 1392 applications. Both upload and download are dominated by 20 applications that cover 95\% and 89\% of total upload and download, respectively.
File downloads, Internet browsing and Facebook account for half the download traffic. Facebook, Google services, Instagram and WhatsApp account for 68\% of upload traffic.

\subsubsection{Data used in this study}
\begin{figure}[tb]
\centering
\includegraphics[scale=0.4]{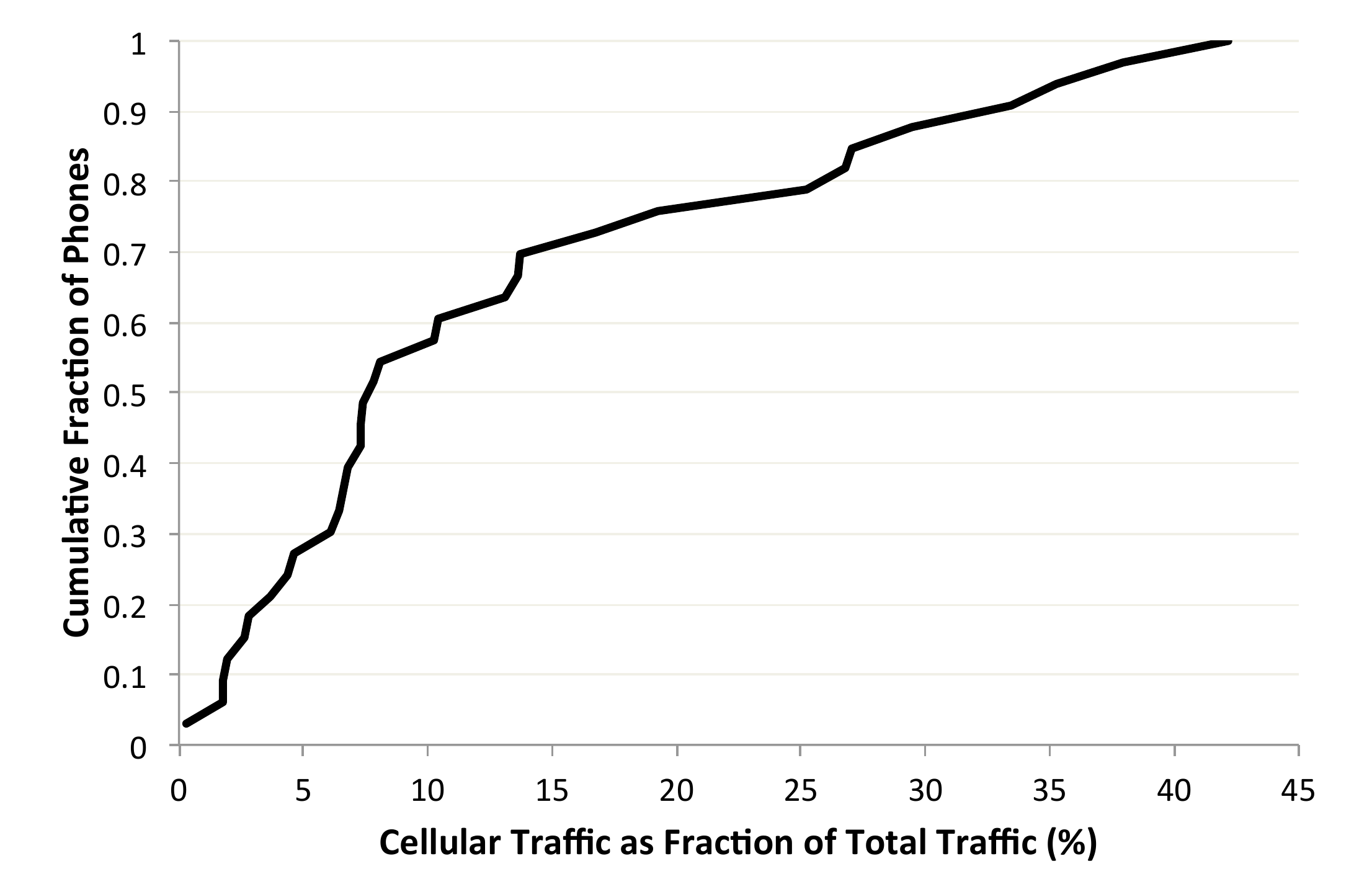}
\caption{Distribution of the cellular traffic as fraction of total traffic.}
\label{fig:percent}
\end{figure}
For this study, we have focused on 45 smartphones, selected because each has contributed at least 5000 measurements (i.e. $\sim$17 days) since May 2015, capturing a total of 1.635.641 measurements, 
representing over 136k hours of measurements. Fig.~\ref{fig:numdays} shows the cumulative distribution of the data collection period lengths in days for these 45 phones. 
Volunteers have joined and left, explaining the variability in terms of collection duration. 
Over 40\% of the phones have reported \app data for over 100 days, with some doing so for over 500 days. 
\begin{figure}[bt]
\centering
\includegraphics[scale=0.4]{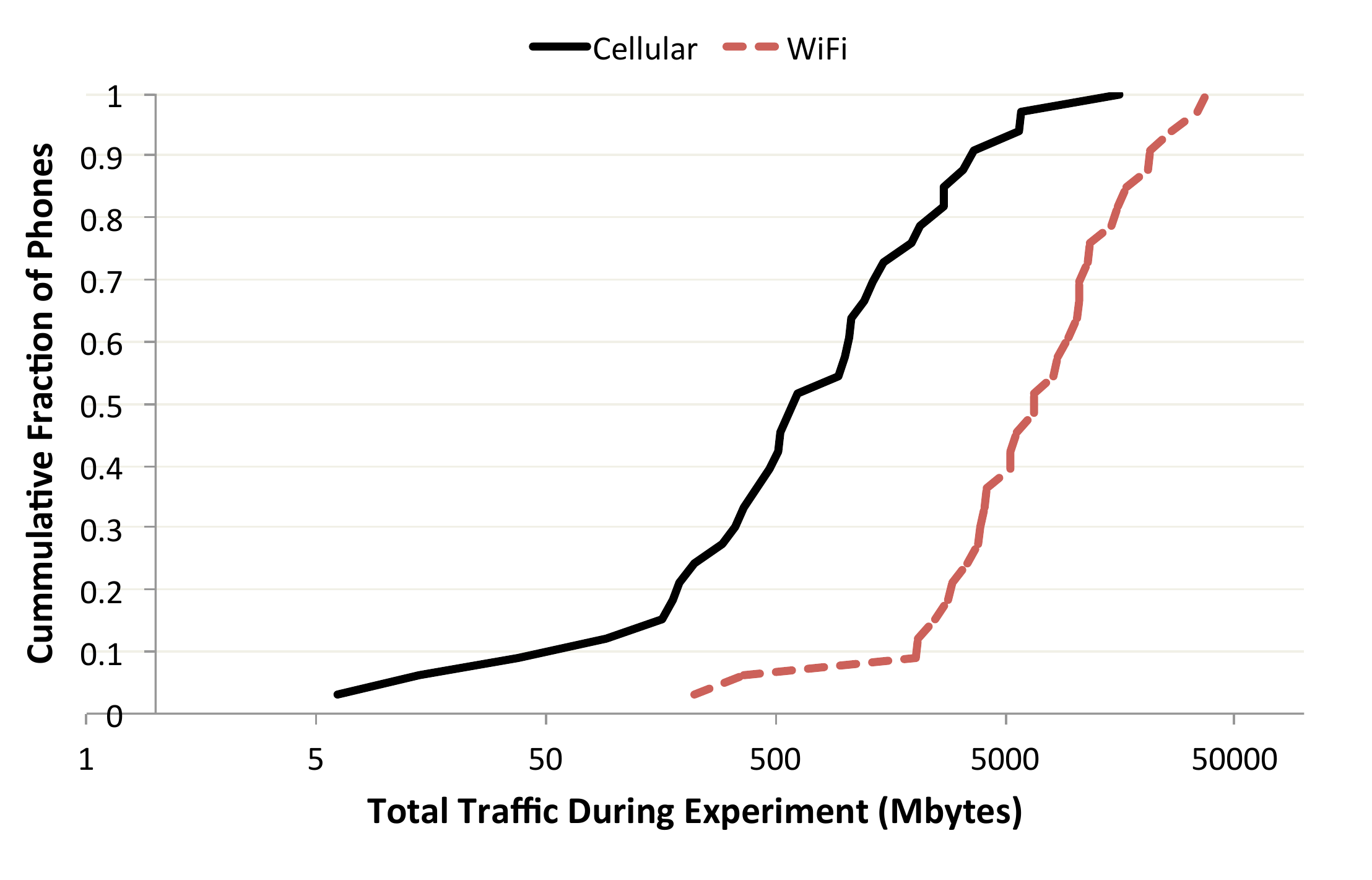}
\includegraphics[scale=0.4]{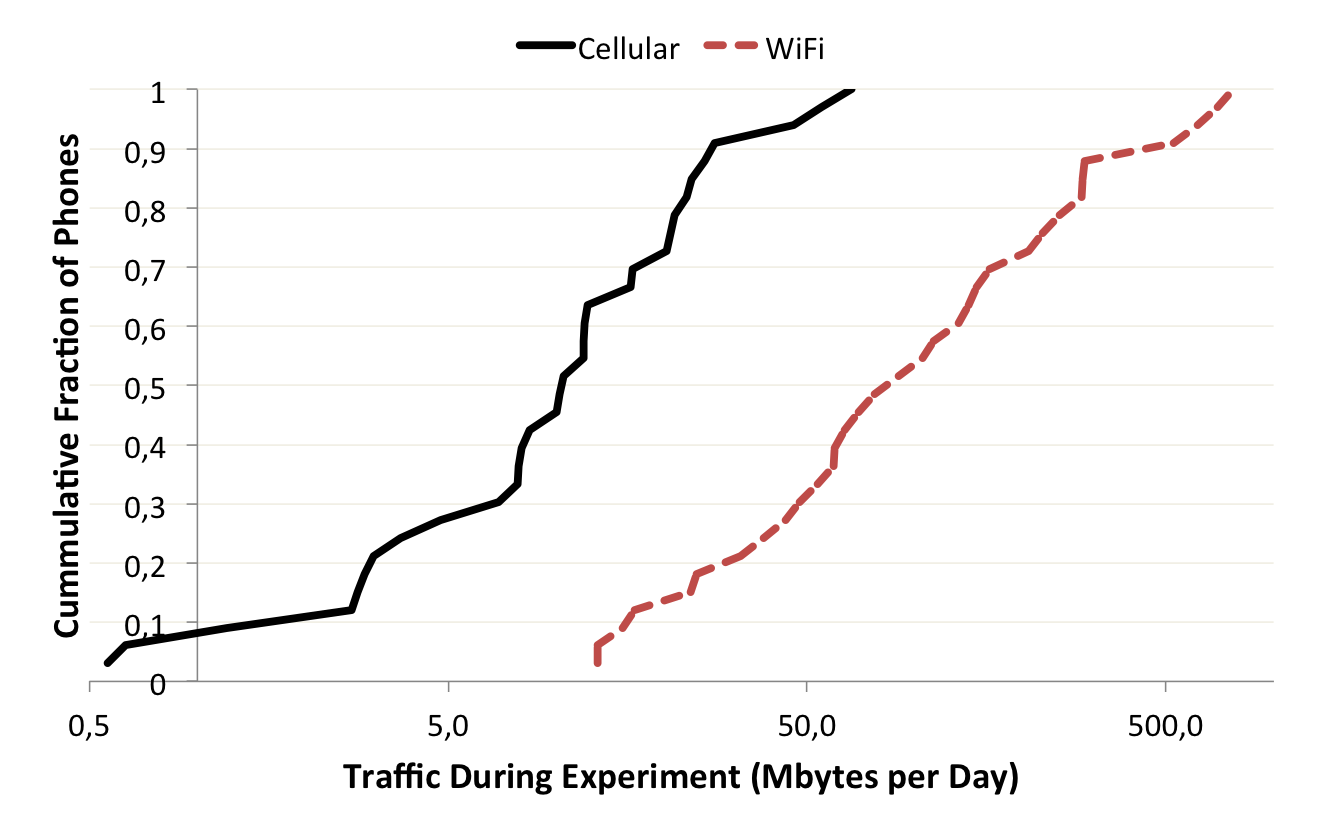}
\caption{Distribution of cellular and WiFi traffic in Mbytes (top) and Mbytes per day (bottom).}
\label{fig:traffic}
\end{figure}
\FloatBarrier

\subsubsection{Cellular and WiFi Traffic}
\label{subsub:cellwifi}
The total traffic generated by 33 smartphones during the data collection period is plotted in Fig.~\ref{fig:traffic}. 
We have removed the traces of 12 phones as these volunteers have generated too little Internet traffic to be meaningful to this study.
Traffic volume is divided between traffic delivered through cellular and that delivered through WiFi. Both upload and download traffic are merged in these statistics. In the data presented here, 
the download volume is 4.26 times the upload volume.  
This data is generated using the following assumptions:
\begin{itemize} 
\item If at a given measurement point the active connection is cellular (resp. WiFi), we assume that the data uploaded and downloaded by all applications over the respective measurement 
period is delivered through a cellular (resp. WiFi) connection. As the measurement period is only 5 minutes, we consider this a fair assumption.   
\item If the data has been delivered through a cellular network but \app
has found in the list of the available WiFi networks the name of a network the user has connected to in the past, we count the data as delivered through WiFi. 
\end{itemize}
The latter assumption is made to analyze fairly the pre-caching opportunities available to all smartphone users. 
Indeed, some users may turn off their WiFi  interface for some time, sending data over cellular when offloading to WiFi might be possible, while other users keep it on all the time. 
To identify all opportunities to offload data over WiFi for all phones, we have decided to change the actual connectivity timeline with this second assumption. 
This modification of the connection timeline has increased the proportion of WiFi traffic significantly for some smartphones. 
In the rest of the paper, we consider an active connection to WiFi following this modified timeline.

\begin{figure*}[htb]
\centering
\includegraphics[scale=0.3]{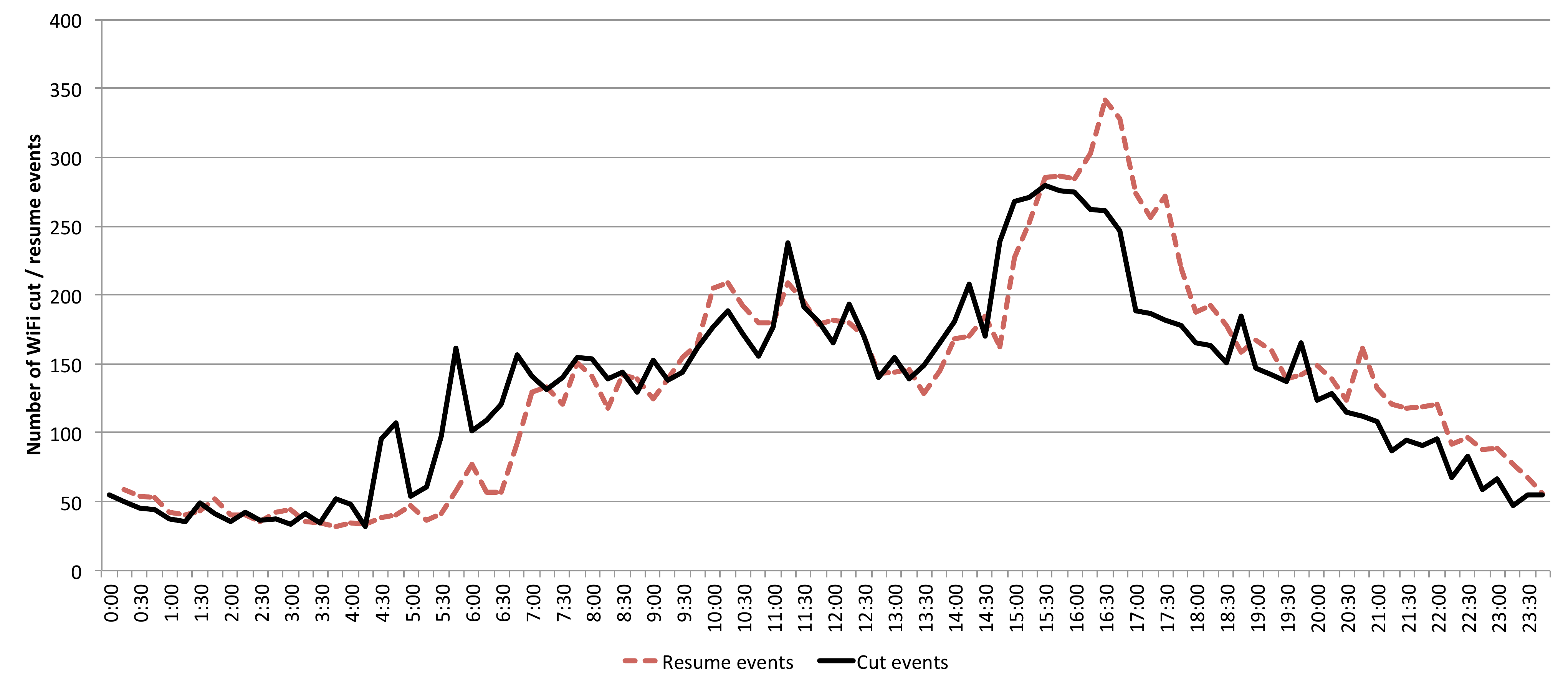}
\caption{Number of WiFi cut and resume events over time using 15-minute time slots.}
\label{fig:CutsResumes}
\end{figure*}

\begin{figure}[tb]
\centering
\includegraphics[scale=0.3]{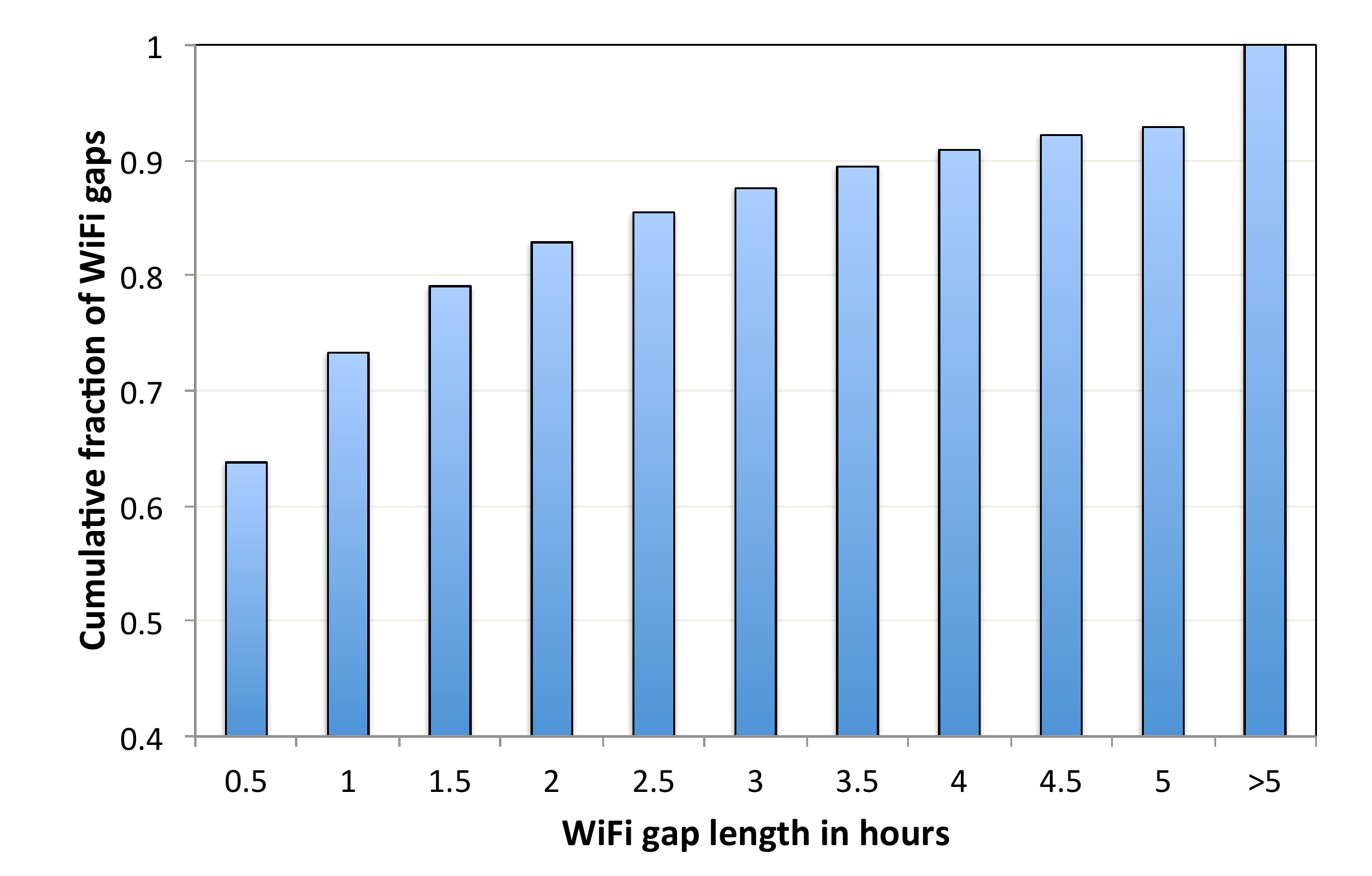}
\caption{Cumulative distribution of WiFi gap lengths.}
\label{fig:WiFiCutDurations}
\end{figure}

\FloatBarrier
\section{Mining for \algo}
\label{sec:mining}

In this section, using the dataset collected by \app, we address what we consider to
be the key questions regarding the feasibility of pre-caching as a strategy for 
reducing peak-hour congestion in cellular networks. Namely, how much traffic
is delivered through cellular networks, what are the gaps in WiFi connectivity that would
justify pre-caching, and finally, what is the ceiling of pre-caching.

\subsection{How much user traffic is delivered through cellular networks?}

Anecdotal evidence and thorough studies~\cite{cisco-vni} show that thanks to the proliferation of
WiFi a significant part of smartphone traffic already flows through WiFi. Therefore,
the first question facing PCach is if enough traffic is still delivered through the cellular network.

Fig.~\ref{fig:traffic}-(top) depicts the traffic volume distribution in Mbytes generated by all the
applications and delivered either through WiFi or the cellular network. 
Fig.~\ref{fig:traffic}-(bottom) depicts the same data in Mbytes per day. 
The data leads to some interesting observations: First, user traffic is highly variable, with some users generating no traffic at all while others generating 
several dozens of Mbytes per day. This is the well known phenomenon of power users already described in \cite{Paul11}. Second, WiFi traffic is an order of magnitude higher than the cellular traffic. 
While the median user generates around 5 Mbytes per day on the cellular network, it generates around 50 Mbytes per day on WiFi, a 10-fold increase.
In total, the 33 phones have generated 56,844 Gbytes over cellular and 329,454 Gbytes over WiFi, cellular traffic representing in this case 15\% of the total traffic. 
This percentage varies over the phones, as shown in Fig.~\ref{fig:percent}. 
If the median user sends 8\% of their traffic to the cellular network, some can send up to 45\% of their traffic. 

In conclusion, our measurements show that, while the proportion of cellular traffic is reduced compared to WiFi, it still represents 15\% of the total mobile data traffic.

The challenge we address later in the paper is whether this 15\% 
of cellular data can be pre-cached, and if yes, is it worth the effort. 

We will show that most probably, this cellular traffic originates from users on the go, usually on their commute.
Commuting times are well known to cause traffic congestion peaks in the cellular networks~\cite{Rossi13}. 
In the subsequent analysis of our fine-grained dataset, we highlight that the gains of pre-caching in terms of traffic volume are quite interesting, but that the most interesting benefit is its ability to reduce peak hour stress on cellular networks.

\subsection{What are the gaps in WiFi connectivity?}

Assuming that the presence of an active WiFi connection absolves the cellular network from having
to deliver smartphone traffic, the next important question facing PCach is if there are gaps in the WiFi connectivity 
and what is their  distribution.

We define a \emph{WiFi gap} as the time period  between a WiFi cut event and a WiFi resume event. 
A WiFi cut event is identified on measurement sample $x$ if:
\begin{itemize}
\item WiFi is the active connection for sample $x-1$,
\item cellular is the active connection for sample $x$,
\item the time elapsed between $x$ and $x-1$ is no longer than 10 minutes.
\end{itemize} 
Conversely, a WiFi resume event is identified on measurement sample $x$ if the previous sample had a cellular active connection while the current sample shows a WiFi connection.
The constraint on the duration in-between timestamps of $x$ and $x-1$ has been added to avoid the negative impact of users disabling \app for an extended period of time.

As mentioned in \cref{subsub:cellwifi}, we consider the active connection to be cellular in sample $x$ 
if the active connection is recorded as cellular \emph{and} no preferred\footnote{A WiFi network the user has connected to in the past.} WiFi network is present in the list of the 
available WiFi networks. Otherwise, it is considered as WiFi.

Fig.~\ref{fig:WiFiCutDurations} and Fig.~\ref{fig:CutsResumes} give relevant statistics on the WiFi  gap periods.
Figure~\ref{fig:WiFiCutDurations} shows the cumulative distribution of the WiFi gap durations\footnote{We have logged the gap periods whose duration is lower than one day.}
and Fig.~\ref{fig:CutsResumes} counts the number of WiFi cut and resume events per 15 minute time intervals of the day. 
The data shows that up to 80\% of the WiFi gaps last no more than one hour and a half while 65\% of them last at most 30 minutes. Only 10\% last more than four hours. 
We can conclude that WiFi disconnection over time is relatively short and within the lifespan of an hour. 

By taking a closer look at the WiFi cut and resume events distribution over time in Fig.~\ref{fig:CutsResumes}, we notice a surge in cuts from 6:00 to 7:00 in the morning 
and from 15:00 to 16:30 in the afternoon. These peaks in cut events are followed, around an hour later, by a surge in WiFi resume events from 9:00 to 10:00 in the morning and from 16:30 to 17:30 in the evening.
 This data clearly follows the patterns of the commuting hours.  
 It is striking how the afternoon interval between WiFi cut and resume events matches the peak-hour cellular traffic data reported in~\cite{Rossi13}, 
which occurs from 15:00 to 16:30. 

The statistics on the WiFi gap duration and its occurrence throughout the day suggest that WiFi disconnections really do occur at commute times. 
Therefore, if it were possible to pre-cache the data users are going to need during their commute, it would be possible to smoothen the peak-hour traffic flowing through the cellular networks. 

\subsection{What is the potential of pre-caching?}
\label{subsec:bound}

\begin{figure}[tb]
\centering
\includegraphics[scale=0.5]{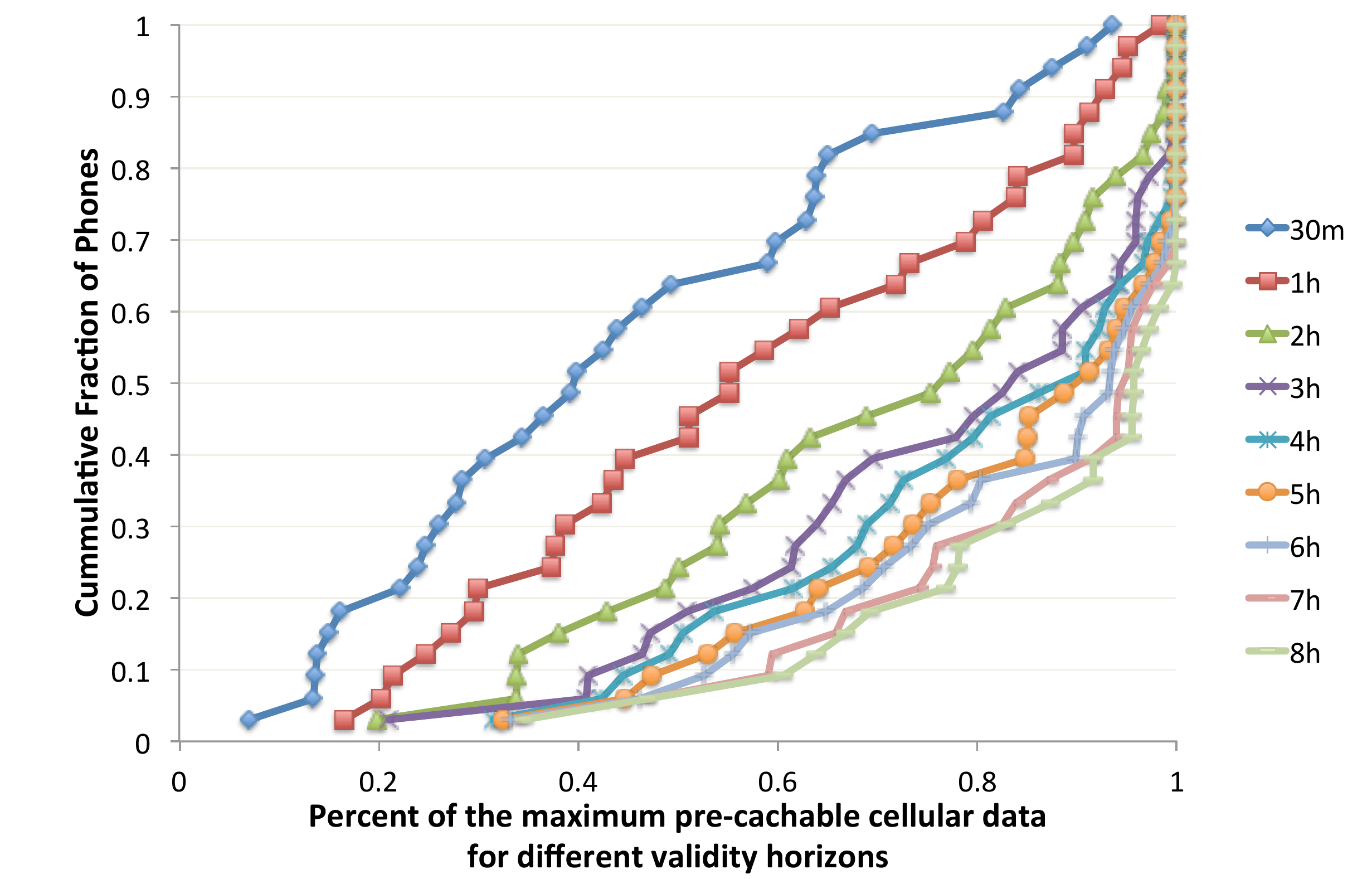}
\caption{Percentage of cellular data that can be pre-cached for different validity horizons (in hours).}
\label{fig:bound}
\end{figure}

We have demonstrated that the occurrence of WiFi gap periods is tightly related to the peak hours of cellular data demand. 
Pre-caching on the smartphone the content users will consume during WiFi gap periods may offload a given volume of cellular data to WiFi. 
If we could perfectly predict the data smartphone users will download over cellular, we would pre-cache it in its entirety 
using the WiFi connection. In this case, the maximum amount of data that can be pre-cached is the cellular data shown in Fig.~\ref{fig:traffic} and Fig.~\ref{fig:percent}, which represents for 
the 33 phones 56,844 Gbytes or 15\% of the total mobile traffic. However, this is a potentially loose upper bound.
In this subsection, we aim at better capturing the potential gains of pre-caching. 

To be able to pre-cache data, it is necessary to predict $i)$ the occurrence of a WiFi gap and $ii)$ the content the user will need during the particular WiFi gap. Assuming we are able to perfectly predict these elements (the predictability of this data is discussed in \cref{sec:pred}), the data pre-cached via WiFi just before a WiFi gap occurs. 
This pre-cached data is relevant at the date of download, but as time elapses, it may get outdated. This would be the case if a newspaper application gets pre-cached and its content is modified during the WiFi gap. 
On the flip side, new content may be generated during the WiFi gap, which, by definition, cannot be pre-cached. 
This is the typical case of emails that are downloaded before the user gets out of its office: it cannot encompass the emails that enter his inbox later.

To capture the time-dependent relevance of data on the potential of pre-caching, we have defined a data validity period called the \emph{horizon}. 
The horizon, expressed in minutes, represents the time for which the pre-cached data is meaningful to the user experiencing a WiFi gap.
 For instance, if we set the value of the horizon to one hour, we assume that the pre-cached data is relevant to the user for at most one hour after it has lost WiFi connectivity.   

For the 33 users in our dataset, we have calculated the amount of \emph{relevant} data that can be pre-cached for different horizon values. 
At each WiFi cut event, we have accumulated the traffic sent over cellular for at most the duration of the horizon or until WiFi resumes. 
For instance, for a horizon of one hour, if the WiFi gap is shorter than one hour, we count all cellular traffic as pre-cached traffic. If the WiFi gap is longer than one hour, 
we just count the cellular traffic transferred for one hour after the WiFi cut time.

Fig.~\ref{fig:bound} shows the fraction of cellular data that can be pre-fetched for different horizon values. The y-axis gives the cumulative distribution of the phones that reach this fraction of cellular data. 
The larger the horizon, the larger the proportion of relevant cellular data we can pre-cache.  
Since we have seen that around 80\% of the WiFi gaps last less than one hour and a half, with a 2-hour horizon we can pre-cache 80\% of the cellular data. 
A 2-hour data validity horizon is reasonable in our opinion. As such, the expected gains of a pre-caching strategy are in the order of 12\% of the total mobile traffic.

\paragraph*{{\bf Study Conclusion}} 
The data analysis in this section shows the potential of pre-caching while providing guidelines on how to design and execute it. The measurements show that a significant percentage  
of mobile traffic is delivered through the cellular network and it is mostly concentrated early in the morning and afternoon, which coincides with commuting hours. Assuming perfect prediction
capabilities, the data shows that up to 80\% of the cellular traffic can be pre-cached, provided users accept that some data can be up to two hours old. Finally, our analysis points
to a pre-caching approach concentrated on reducing peak-hour traffic congestion in cellular networks as the most promising strategy. 



\section{PCach: A user-centered pre-caching strategy}
\label{sec:pred}

In this section, we introduce \algo, a pre-caching strategy whose design is driven by the analysis of the measurements dataset.  
It consists of predicting the future occurrence of a WiFi gap and the data a user is likely to need during the gap so as
to be pre-cached on the smart device.
This approach concentrates on predicting the future occurrence of a WiFi gap. 
If a WiFi gap is predicted to happen, \algo selects a subset of applications for which it is beneficial to pre-cache data. 
Such applications are the ones that are highly likely to download data from the Internet during the WiFi gap. 
In the following, we describe how \algo is structured and how we address the challenges of predicting the WiFi gaps and 
what content is beneficial to pre-cache.

\subsection{The \algo approach}

\algo can be implemented as a standalone mobile application. 
Thus, it only leverages information that can be accessed through the native Application Programming Interface (API) of the operating system. 
As a main goal of our design is protecting user privacy, all the exploitation and storage of sensing data for prediction purposes is performed exclusively on the smartphone.

\begin{algorithm}[!htb]

\LinesNumbered
\SetKwInOut{Input}{Input}
\SetKwInOut{Output}{Output}
\SetKwComment{Comment}{//}{}
\SetNlSty{textbf}{}{ :}
\small{
\Input{current time slot, $cSlt$; list of pre-cachable apps, $sApps$; number of apps to pre-cache, $K$}
\Output{List of apps to pre-cache, $PCachApps$}
\Begin{
{\bf PastDB} $\leftarrow$ Update history($cSlt$) \;
\Comment{\textcolor{ForestGreen}{Is there a WiFi gap in the next slot?}}
$cut \leftarrow$ predictNextWiFiCut({\bf PastDB}, $cSlt$)\;
\If{cut $==$ true}{
    \Comment{\textcolor{ForestGreen}{When will WiFi resume?}}
    $rSlt \leftarrow$ predictWiFiResumeSlot({\bf PastDB}, $cSlt$)\;
    \Comment{\textcolor{ForestGreen}{Get list of top apps during WiFi gap}}
    $PCachApps \leftarrow$ predictTopKApps($sApps$, $K$, $cSlt+1$, $rSlt$)\;    
}

Return $PCachApps$;
}
}
\caption{{\bf PCACH}}
\label{algo:PCach}
\end{algorithm}

Once the \algo app is installed on a smartphone, it triggers Algorithm~\ref{algo:PCach} periodically. 
To predict WiFi gaps, \algo divides a 24-hour period into time slots. In any given time slot, $cSlt$,
it first predicts if a WiFi cut event is going to occur in the next time slot, $cSlt +1$. If
that is the case,  it predicts the identifier $rSlt$ of the future slot where WiFi is supposed to resume, giving
\algo the information necessary to know when a WiFi gap occurs and its duration.  
If the duration is non-zero, it predicts the top $K$ from a list of applications, $sApps$, considered to be \emph{pre-cachable} (more on this in~\cref{subsec:appPred}).

All predictions rely on the smartphone usage history, stored in a local database, $PastDB$,
which is updated continuously. 
It stores the following data features:
\begin{itemize}
\item WiFi scan : the list of visible WiFi networks.
\item Whether the mobile phone's active network is WiFi or cellular. If its active connection is WiFi, it records its network name.
\item The list of applications currently running,
\item The volume of data that has been uploaded and downloaded per application to the Internet since the last {\bf PCACH} algorithm call.
\end{itemize}

The efficiency of \algo obviously depends on the efficiency of the algorithms used for predicting WiFi gaps and which applications to pre-cache. 
They have to be light enough in terms of processing and only use local data to ensure privacy. 
Next, we introduce simple prediction strategies, either based on statistics or on state-of-the-art machine learning algorithms, and show that they perform well on our dataset.

In terms of implementability, the P-Cach app derives from our crowdsensing app as it measures periodically the same data as the one in our data set. This measurement step is easily done in a standalone app. To trigger prefetching for a given app, P-Cach simply requests the operating system to launch the app for example by pushing it to the foreground for a short period of time. The app is thus likely to update its content in a standalone manner. The consequence is that prefetching only happens if the app is configured to update its content in running status. 

In terms of memory, the context and content data we store is really light: a day of measurement generates 158 bytes of raw data, in average. This can be further reduced in volume as our prediction schemes build on very simple quantitative information extracted from these measurements (cf. Table IV) at each measurement date. 
The prefetched content will add to it, but since we let the app prefetch the content we rely on its own memory management policy. Moreover, compared to the couple of gigabytes of memory available in today's smartphones, the prefetching impact on memory is limited and for us, not a critical problem. 

In terms of energy footprint, it is the periodic data collection that is energy-hungry. P-Cach uses the same data collection operations as our crowdsensing app. This app has been designed with care to minimize energy consumption. Main design choices are clearly explained in [10]. With a data collection period of 5 minutes, a regular Android phone operates 2 to 3 days long on battery and even when all types of measurements listed in Table I are made in parallel. For P-Cach, the data collection is made every 15 minutes, and most energy-hungry measurements such as GPS, bluetooth or accelerometer are not necessary. Prediction and prefetching tasks impact less the overall energy consumption.

Next, we address the problem of predicting the $K$ applications that are the most likely to transfer data within a given time interval, as required for the instruction 6 of Algorithm~\ref{algo:PCach}. Second, we address the WiFi gap prediction, highlighting how the recorded history has to be handled to offer the best possible prediction performance in terms of sensitivity and specificity. 

\subsection{Top app prediction} 
\label{subsec:appPred}

The goal of this prediction is to select the top $K$ applications most likely to send or receive data during the WiFi gap.
This problem is closely related to the one addressed in \cite{Parate2013} where authors aim at predicting the mobile application that is the most likely to be used next.
Since the app that is the most likely to send or receive data is related to the application usage, it is possible to leverage the prediction algorithms of the literature such 
as Prediction by Partial Match (PPM) \cite{ppm-1984} or N-grams \cite{ngram}. However, in our context, we have to carefully choose the set of applications that are 
relevant to pre-cache for the user by favoring the ones whose data is relevant during the WiFi gap. Furthermore, we also have to decide on the appropriate 
value of $K$, the number of apps to pre-cache. 

In the following, we extract from our dataset meaningful hints as to the choice of $K$, 
the set of pre-cachable apps, $sApps$, and the algorithm to predict the top $K$ apps to pre-cache.  

\subsubsection{Choice of $K$}
\begin{figure}[t]
\centering
\includegraphics[scale=.5]{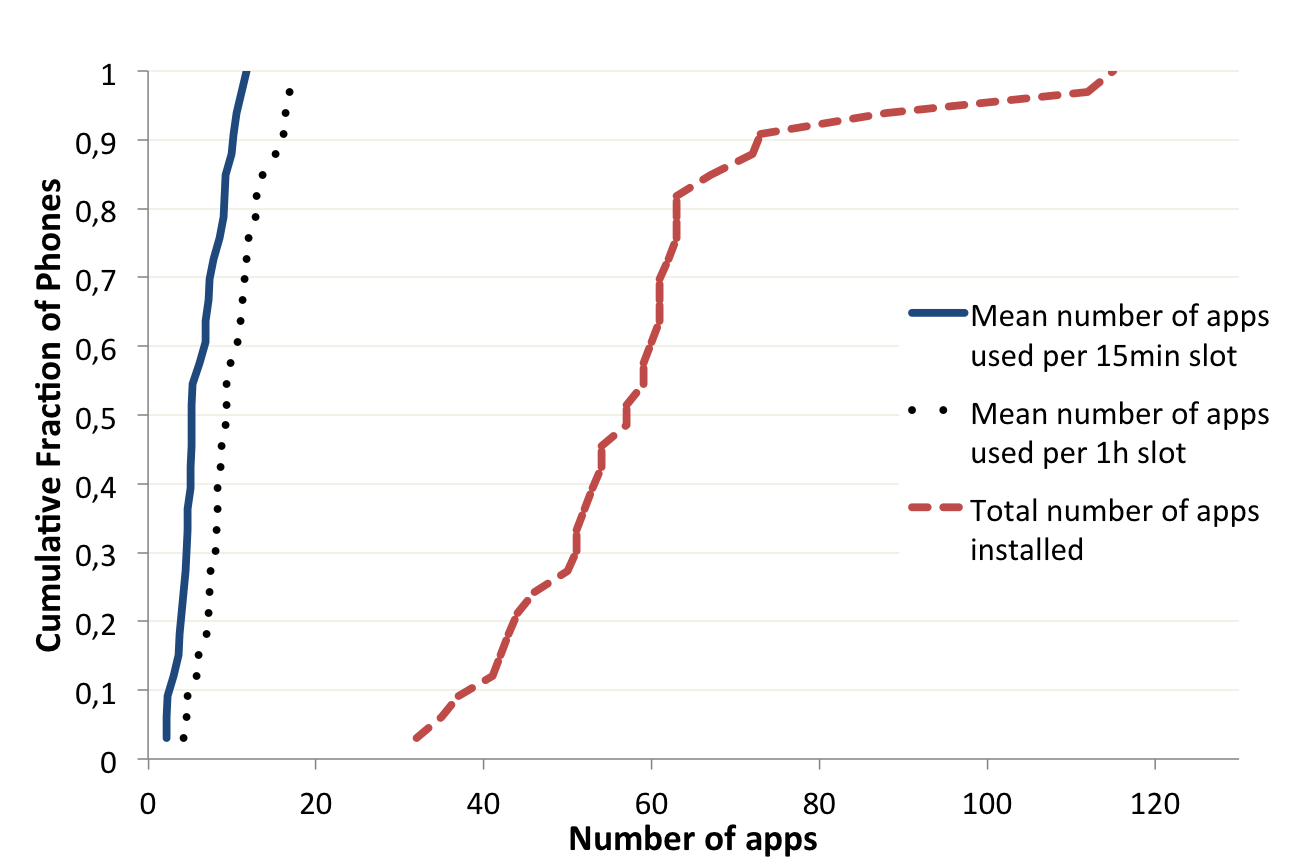}
\caption{Cumulative distribution of the average number of applications that create mobile traffic per 15 minutes slot over 33 phones. It is compared to the total number of apps installed.}
\label{fig:MeanAppNb}
\end{figure}

Figure~\ref{fig:pieApps} shows that few apps trigger most of the download data, thus we are looking for a small value of $K$.
Figure~\ref{fig:MeanAppNb} plots the cumulative distribution of the average number of applications used per phone per time slot, for slots durations of 15 minutes and 1 hour.
The median phone only runs about 5 apps (resp. 10 for a 1h slot) which trigger data transfers, an order of magnitude lower than the number of apps installed.
For the 1-hour slot, the increase is limited as the median phones use only about 10 apps, all values varying between 4 and 17.    
As a result, it seems reasonable to set $K \leq 10$ for the 15-minute slot and $K \leq 17 $ for the 1h-long slot. 

\FloatBarrier

\subsubsection{Pre-cachable application set}
Among the set of applications that send or receive data, we have identified categories of applications that have good properties in terms of data validity. 
We say that these applications are \emph{PCachable}. Regular messaging applications, downloads, Internet browser, radio or music streaming applications are typically not PCachable.
 On the other hand, social media apps such as Facebook, Tweeter or Instagram are PCachable as it is possible to pre-cache the actual status of the list of news (i.e. the current wall) available.
 Of course, news, weather forecast and sports apps are PCachable as well. We have also included video streaming apps such as YouTube or Twitch as they have personalized channels
 whose latest data can be pre-cached. 
 Finally, Maps-like applications can be updated with local map information at pre-caching date. 
Table~\ref{tab:pcachable} lists, for the 20 top applications represented in Fig~\ref{fig:pieApps}, whether we consider them ``PCachable", the percentage of total traffic they represent and the percentage of 
measurements in which they appear on the list of the recently used apps.
The table is ordered by decreasing order of the last column.

The PCachable applications account for a total of about 29\% of the total traffic. 
As such, with a back-of-the-envelope calculation, and knowing that the potential gain for the median 
user for a horizon of 2 hours with PCach was of 12\% of the total mobile traffic, we get a gain of 4\% of 
total traffic with the PCach approach. This amount can be really beneficial for the cellular network as it will particularly reduce peak hour traffic demand.

\begin{table}[t]
\centering
\caption{Top 20 PCachable applications in our dataset}
\label{tab:pcachable}
\begin{tabular}{c|l|c|c}
PCachable & Application & \begin{tabular}[c]{@{}l@{}}Percent of \\ total traffic\end{tabular} & \begin{tabular}[c]{@{}l@{}}Percent of \\ Appearance\end{tabular} \\ \hline
0 & Other apps & 7,84\% & 35,073\% \\
0 & Google + Phone services & 2,15\% & 31,882\% \\
0 & WhatsApp & 1,69\% & 7,791\% \\
0 & Internet browser & 9,21\% & 7,749\% \\
\textbf{1} & Facebook & 14,01\% & 6,148\% \\
\textbf{1} & E-mail & 0,76\% & 3,294\% \\
\textbf{1} & Maps & 1,28\% & 2,182\% \\
\textbf{1} & Instagram & 4,79\% & 1,994\% \\
\textbf{1} & News apps & 0,40\% & 1,478\% \\
\textbf{1} & YouTube & 2,80\% & 0,649\% \\
0 & Downloads & 16,74\% & 0,356\% \\
{\bf 1} & Sports apps & 1,29\% & 0,295\% \\
0 & Spotify & 0,75\% & 0,242\% \\
{\bf 1} & 9GAG & 1,42\% & 0,200\% \\
{\bf 1} & Twitter & 0,41\% & 0,199\% \\
0 & Snapchat & 2,06\% & 0,166\% \\
0 & Netflix & 1,27\% & 0,132\% \\
0 & Deezer & 1,16\% & 0,089\% \\
{\bf 1} & Twitch & 1,83\% & 0,065\% \\
0 & TuneIn Radio & 0,40\% & 0,040\% \\
0 & TRENDnetVIEW & 0,34\% & 0,002\%
\end{tabular}
\end{table}

\subsubsection{History-based prediction}


The next step is to discuss how to actually predict the top $K$ apps.  Table~\ref{tab:pcachable} shows that there is a strong correlation between the number of times used  and the proportion of data 
being sent over the Internet. This is not surprising but it gives us a very simple feature that can be used to predict the top $K$ apps to be pre-cached: the number of times the app has been called.  

We have implemented a basic prediction algorithm that creates, for each PCachable app, a histogram $H$ that counts the number of times an app has been called in a specific time slot. 
The top $K$ apps are simply those with the highest histogram values.
We simply rank the PCachable apps by decreasing order of their occurrence counted in the current slot $cSlot$ (i.e. $H[cSlot]$) and select the $K$ top ones to pre-cache. 
Histogram is updated at instruction 2 of Algorithm~\ref{algo:PCach} by incrementing by one the histogram element $H[cSlot-1]$ of the previous slot $cSlot-1$ for each PCachable app that has run since last {\bf PCACH} run.
At instruction 6 of Algorithm~\ref{algo:PCach}, we simply rank the PCachable apps by decreasing order of their occurrence counted in the current slot $cSlot$ (i.e. $H[cSlot]$) and select the $K$ top ones to pre-cache. 

This prediction algorithm is tailored for a prediction horizon of one time slot (i.e. 15 minutes in our case). It can be easily adapted to a WiFi gap duration larger than one slot by iterating the same top $K$ app selection for each slot of the WiFi gap, and prefetching the union of all sets of $K$ apps predicted. 
For instance, it the WiFi gap lasts 3 slots, the union of all 3 sets of $K$ apps will be pre-fetched. 

\subsubsection{Results}
\label{subsub:historyresults}
We have tested our simple light-weight app-prediction algorithm on our dataset. Since for each slot we know exactly the set of apps that have actually sent data over the cellular connection,
 we can compare our predictions to the ground truth. For each phone, we have used a week for initial training of the histogram. 
We have varied the value $K$ and calculated for each phone the true positive rate (i.e. sensitivity) and the false positive rate (i.e. 1-specificity).
The true positive rate is defined as $TPR = TP / (TP+FN)$, with TP the number of true positives and FN of false negatives. The false positive rate is defined as $FPR = FP / (FP + TN)$, 
with FP the number of false positives and TN of true negatives. A perfect prediction algorithm offers a $TPR$ equal to 1 and a $FPR$ equal to 0. 

\begin{figure}[tb]
\centering
\includegraphics[scale=0.5]{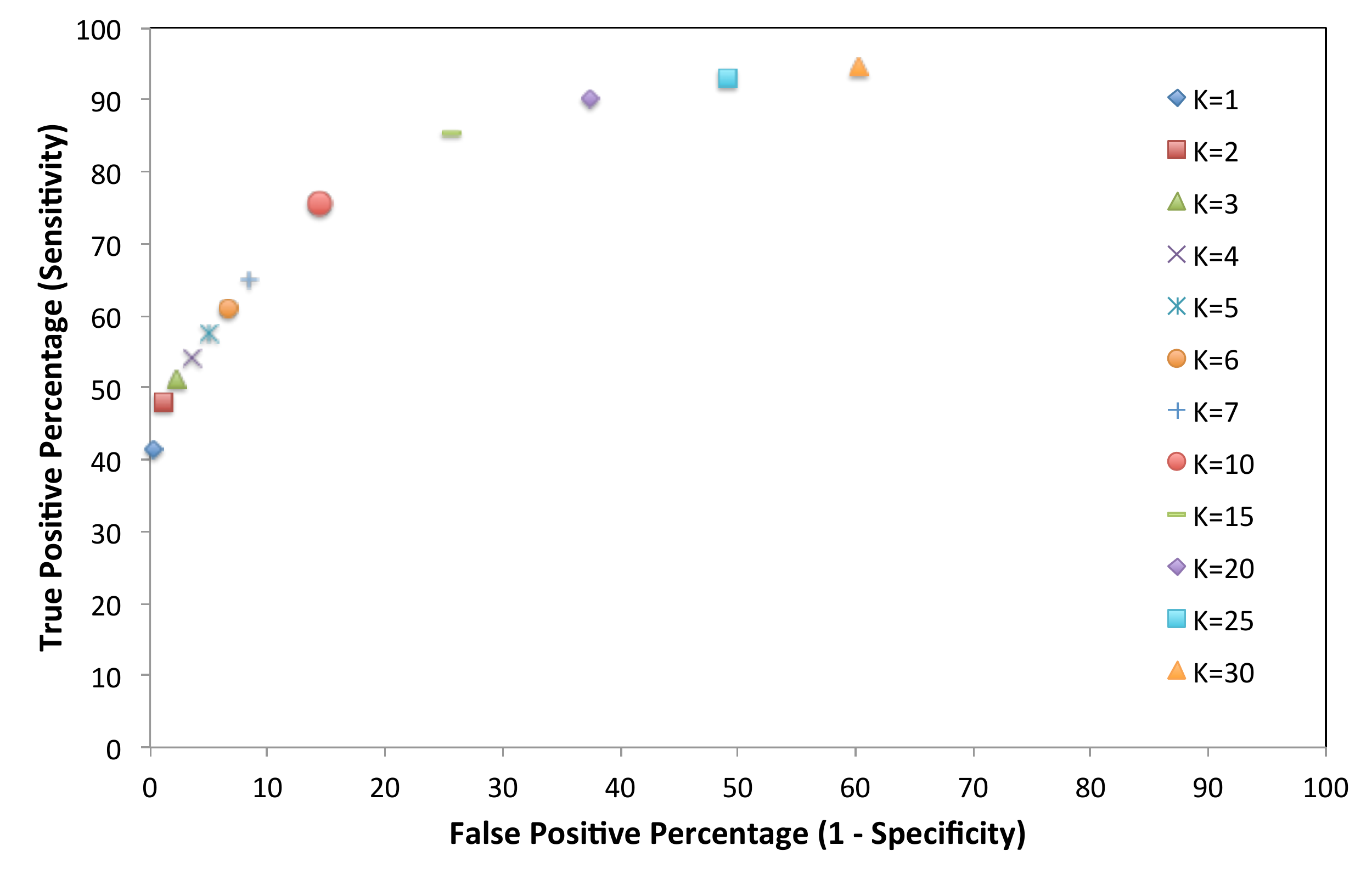}
\caption{Mean percentage of true positives as a function of false positives for $K$ top application prediction. Each point represents the average of TPR and FPR over all phones for different $K$ values.}
\label{fig:ROCapp}
\end{figure}

In our case, a true positive occurs if the predicted app is used during the WiFi gap. A false positive occurs if a predicted app is not used during the WiFi gap.
A false negative is counted if there is an app actually used that is not predicted. And finally, we have as many true negatives as PCachable apps that haven't been selected in the top $K$ apps.   
Figure~\ref{fig:ROCapp} shows the performance of our algorithm for values of $K=\{1,2,3,4,5,6,7,10,15,20,25,30\}$. It represents the $TPR$ as a function of the $FPR$, given in percents.
 Optimal prediction is represented by the Oracle point at coordinates (0,100).
Each point represents the average (TPR, FPR) calculated over all phones for a different value of $K$. As $K$ increases, the TPR increases but at the cost of more false positives. False positives would trigger
unnecessary pre-caching and, as such, it is reasonable to reduce them. The data shows that a good compromise is to select a $K$ value lower than 10. 
In this case, the overall prediction quality is very good with less than 20\% false positive rate. 
A false positive rate of 20\% means that about 20\% of the installed apps that trigger traffic are pre-cached unnecessarily, which represents less than 8 apps for the median user.

\begin{figure}[bt]
\centering
\includegraphics[scale=0.5]{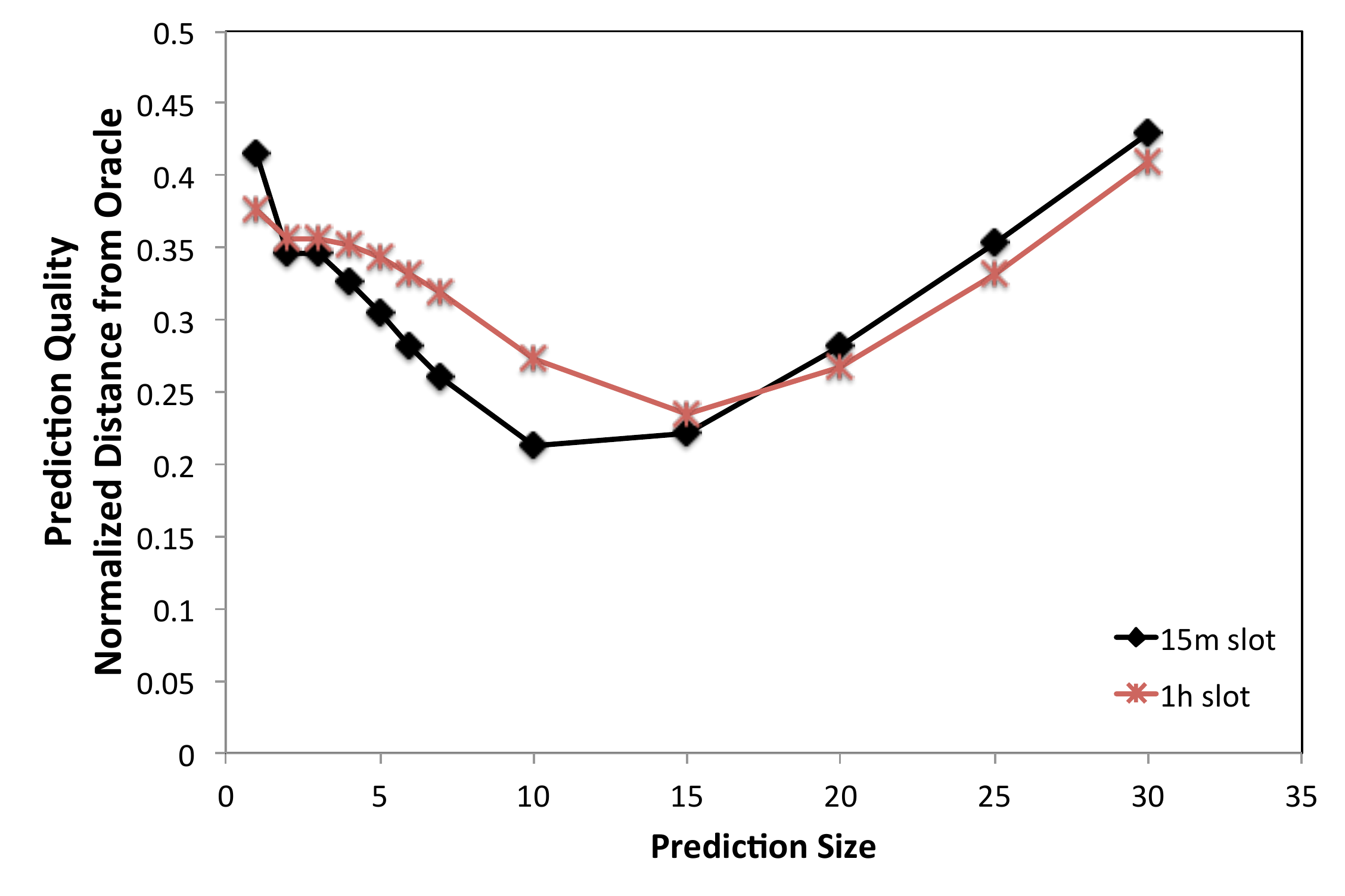}
\caption{App prediction quality gap as a function of $K$. Prediction quality is measured by the normalized distance to the Oracle.}
\label{fig:K}
\end{figure}

As the WiFi gap can last more than 15 minutes, Fig.~\ref{fig:K} shows the prediction-quality gap as function of $K$ when using 15-minute and 1h slots. The prediction-quality gap is measured by calculating 
for each point of Fig.~\ref{fig:ROCapp} its Euclidean distance to the Oracle point, normalized to one. The closer this prediction metric is to 0, the better the respective prediction.
  This metric has been plotted as a function of $K$ in Fig.~\ref{fig:K} for 15 minutes and 1 hour slots. 
  For 1 hour slots, the $K=15$ triggers a false positive rate equal to 20\%, which is reasonable. 
The data shows that for 15-minute slots, setting $K$ to $10$ yields the best prediction while for 1h slots this result is achieved for $K=15$.
 Both values are in the order of the average number of apps creating traffic for both slot size values. Indeed, based on Fig.~\ref{fig:MeanAppNb}, we had conjectured that 
 values of $K$ around $10$ and $15$ for the 15-minute and 1h slots, respectively, looked most likely to be the right choices. 

From these first developments, we can conclude that it is possible with a simple history-based prediction algorithm to select the top $K$ apps efficiently. 
We have shown that it is possible to adjust $K$ to the length of the WiFi gap. These initial tests open the way to exciting future research.  
Next, we will look deeper into our data and leverage it to learn the best $K$ for a given user and slot size. 
More sophisticated algorithms can be tested as well, most probably inspired by PPM or n-grams.
\begin{table}[!h]
\centering
\caption{Features implemented for AdaBoost learning. The value for each feature is calculated at each time slot.}
\label{tab:features}
\begin{tabular}{c|l|l}
 & Data type & Description                                                                   \\ \hline
1       & boolean   & Covered by home WiFi network from 8pm to 8am                              \\
2       & boolean   & Covered by work WiFi network from 8am to 8pm                           \\
3       & boolean   & It is a weekday or a weekend day                                              \\
4       & integer   & The number of seen WiFi networks                                              \\
5       & boolean   & Top 1 Preferred WiFi is in the list of seen WiFi                         \\
6       & boolean   & Top 2 Preferred WiFi is in the list of seen WiFi                         \\
7       & boolean   & Top 3 Preferred WiFi is in the list of seen WiFi                         \\
8       & integer   & Index of current slot                                                         \\
9       & float     & Probability of cut / resume for this specific slot 
\end{tabular}
\end{table}
\subsection{WiFi gap prediction} \label{subsec:WiFiPred}

\begin{figure}[bt]
\centering
\includegraphics[scale=0.5]{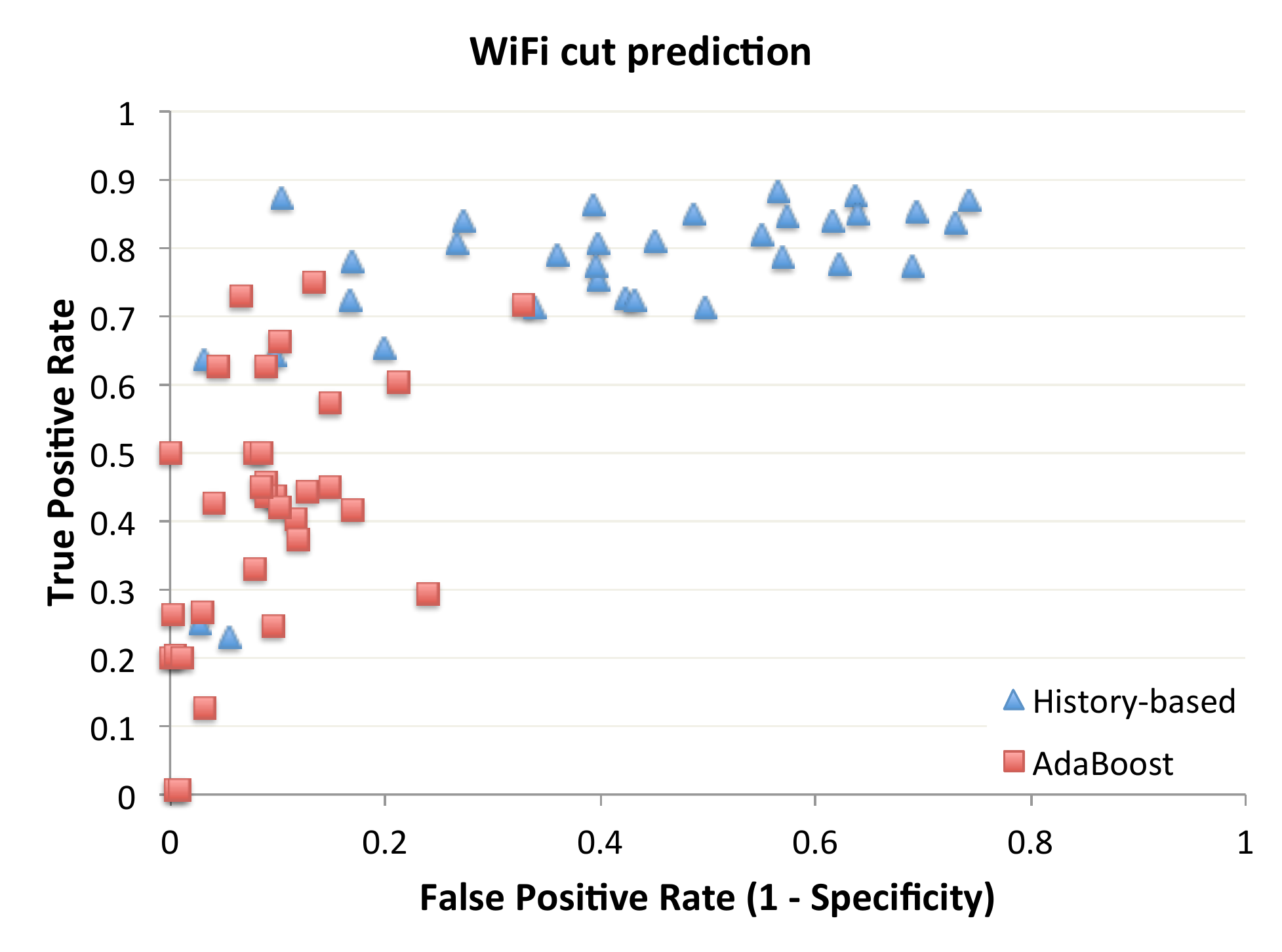}
\includegraphics[scale=0.5]{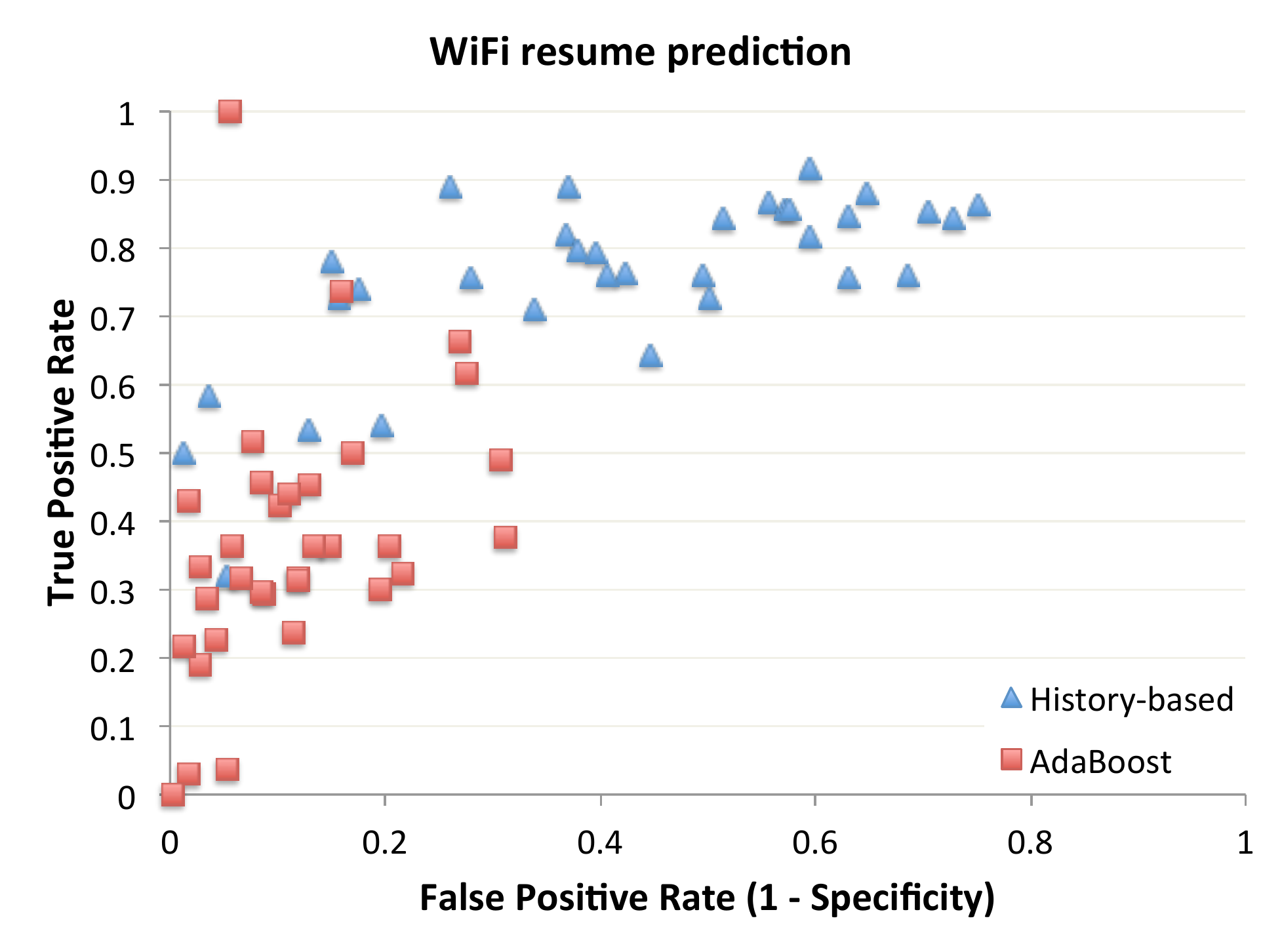}
\caption{WiFi cut (top) and resume (bottom) prediction results: true positive rate as a function of false positive rate. 
Results are given for history-based and AdaBoost learning algorithms.
}
\label{fig:WiFiPred}
\end{figure}

A central step of Algorithm~\ref{algo:PCach} is the WiFi gap length prediction. It is decomposed into two challenges: predicting a WiFi-cut event and predicting the slot in which WiFi resumes. 
Both steps are related to detecting the events presented in Fig.~\ref{fig:CutsResumes}. 
It is very important for the WiFi cut predictions to keep the false positive rate low to limit the number of unnecessary pre-cache operations. 
What makes this criterion particularly important is that our data shows that the proportion of cuts is low compared to the proportion of no-cuts.
For 15-minute slots, on average, only 2\% of slots contain a WiFi-cut event.
As a result, a 40\% false positive rate would predict a wrong cut almost 38 times per day.   
Two prediction algorithms have been tested: The first is similar to the history-based app prediction. The second is based on AdaBoost~\cite{adaboost}, a state-of-the-art machine learning algorithm. 
Fig.~\ref{fig:WiFiPred}  show the prediction results for the WiFi cut (left) and resume (right) prediction, respectively. 
Both plot the TPR as a function of the FPR for both history-based and AdaBoost algorithms. It can be clearly seen that AdaBoost performs much better than the history based prediction as it limits the false positives rate to around 20\%. Implementation details are given next, but the main takeaway is that WiFi cuts are more difficult to predict as the proportion of cuts in reality is really low. AdaBoost is known to better tackle datasets where the proportion of the two

\FloatBarrier

\subsubsection{History-based predictions}
We have developed a history-based algorithm that predicts whether WiFi is likely to be cut or to resume in the next slot.   
This algorithm creates a histogram by counting cuts for all slots of the day. At prediction time, a random value between 0 and 1 is thrown $N$ times. 
We count the number of times $X$ this value is lower than the cut probability $p_{cut}$ of the current slot stored in the histogram. If $X$ belongs to the interval $[(1-\delta)*p_{cut},(1+\delta)*p_{cut}]$, a WiFi cut is predicted. 
Resumes are predicted the same way. Different values of $N$ and $\delta$ have been tested to maximize the prediction quality metric (\cref{subsub:historyresults}). 
The results in Fig.~\ref{fig:WiFiPred} are plotted for $N=10000$ and $\delta=0.1$.

\subsubsection{AdaBoost prediction}

AdaBoost (Adaptive Boosting)~\cite{adaboost}, an ensemble-learning method, is selected as it is considered one of the best off-the-shelf classification methods~\cite{Wu2008}.
To apply it on the WiFi gap prediction,  we used with the set of features listed in Table~\ref{tab:features}. 
The features are a result of our effort to define the user's context based on the WiFi networks seen. For example,
features $5, 6, 7$ are based on how often certain WiFi networks are seen by the smartphone in the slot in which
the prediction takes place.  
They capture relevant information from the list of WiFi networks seen by the phone in the current time slot. 
From this list of WiFi networks a smartphone records, it is possible to extract the subset of networks the user has access to: the preferred WiFi network list. 
From the preferred network list, we extract from the training set the number of times these preferred WiFi networks have been seen. 
Using this count, we extract the  3 most often seen networks and record their names.  
Every time one of these top 3 networks is seen, we can update features 5, 6 and 7 accordingly. 
Features $1, 2$ tell whether the \emph{home WiFi} or the \emph{work WiFi} is currently seen in the preferred WiFi network list, capturing in a simple manner the user location. 
Home and work WiFi are identified by selecting the network seen the most often during the night and during the day, respectively. 
Feature 1 is true if the slot belongs to the nigh time and the home WiFi is seen in the list of the available WiFi networks. Conversely, feature 2 is true if the slot belongs to daytime and the work WiFi is present. 
Training has been performed on a per phone basis, each one using half of its available data. The obtained model has been used for predicting the other half.

The results given in Fig.~\ref{fig:WiFiPred} show that Adaboost is the best solution for PCach: it delivers
a good recall rate with a low false positive rate. 

\section{Related Work}
\label{sec:related}

This paper touches on several subjects, including 5G~\cite{5GBE}, mobile data offloading~\cite{offloading-ton-2013}, mobile traffic analysis~\cite{Paul11}, proactive caching~\cite{Debbah-caching-2014},  
machine learning~\cite{adaboost}, etc. so a thorough presentation of the relevant literature is beyond the scope of this section. Instead, in the following,
we present a few representative works and refer the interested user to the references therein.

As the smartphone becomes our gateway device to the internet, the mobile data is exploding, putting an enormous strain on the cellular
infrastructure~\cite{cisco-vni}. 5G is positioned as the answer but currently it is more of an umbrella of visions, including millimeter Wave (mmWave)~\cite{rangan2014,80211ad}, ultra-densification~\cite{6620224,andrews2012}, 
massive multiple-input, multiple-output (MIMO)~\cite{marzeta2007,marzeta2010} and proactive caching~\cite{Debbah-caching-2014}. All these solutions
require radical changes to the telecommunications infrastructure. The mmWave solutions propose using the 57 GHz -- 64 GHz spectrum, colloquially referred to as
the 60 GHz, which as of now remains mostly idle, for good reasons. It suffers from strong pathloss, atmospheric and rain absorption, poor penetration
of obstacles, etc, making it, traditionally, suitable only for short, static, line-of-sight links~\cite{rappaport2011}. Ultra-densification involves shrinking the cell size to the point of having a base station per client, which can then
fully exploit the former's backhaul connection capacity. However, user association, already a complex
combinatorial problem will explode in complexity once the requirement of a single user per base station
is added. Massive MIMO involves installing potentially hundreds of antenna elements on the base station and a single element on mobile
devices~\cite{marzeta2007}, delivering massive enhancements in spectral efficiency without the need for cell densification~\cite{hoydis2013}. However, it
still remains a theoretical concept, facing many challenges before it can transition into a product. Proactively caching content at base stations is proposed as
a solution for beyond 4G wireless networks~\cite{Debbah-caching-2014}. However, this approach requires significant changes to the base station infrastructure and raises 
several questions regarding subjects such as net neutrality. While transitioning to 5G is inevitable, and will almost certainly involve a combination of the visions discussed, 
we are clearly years away from a large-scale roll-out. PCach, as a light-weight,
out-of-the box solution addresses an immediate need for peak congestion relief.

Exploiting the smartphone's multiple communication interfaces to shift traffic between different networks, also known as offloading~\cite{offloading-ton-2013,offloading-survey} and/or onloading~\cite{Rossi13},
has been proposed to address congestion. Such solutions presume the simultaneous presence of multiple networking technologies, in particular, WiFi and cellular. However, as our data analysis has shown,
for a significant percentage of the time, only the cellular network is available to users.

\section{Conclusions}
In this paper, we followed a progressive and data-driven approach that led to the design of PCach: 
A user-centric solution that proactively caches content users will likely
need during WiFi gap periods so as to reduce congestion in cellular networks. We first built the case
for proactive caching by showing the existence of significant WiFi gaps, in particular at commute times, and cellular-traffic 
only. We then instantiated PCach by introducing an algorithm capable of predicting when and what to cache.  
As future work, we intend to release PCach as an app on Google Play and evaluate its performance on real smartphone
users, using the recruiting mechanism we used for \app and relying on some of the same volunteers.

\section*{Acknowledgment}
This work is supported in part by CHIST-ERA MACACO project, ANR-13-CHR2-0002-06. 
Authors would like to thank Tao Peng and Aiman Elhaimer for their contribution to the early developments of this work.

\end{document}